\documentclass[review]{elsarticle}

\graphicspath{{figures/}}

\usepackage{lineno,hyperref,todonotes,upgreek,amsmath,paralist,subcaption}

\newenvironment{psmallmatrix}
  {\left(\begin{smallmatrix}}
  {\end{smallmatrix}\right)}

\modulolinenumbers[5]

\journal{Journal of Molecular Spectroscopy}

\biboptions{numbers,sort&compress}
\begin{document}

\begin{frontmatter}

\title{Visible and Ultraviolet Laser Spectroscopy of ThF}

%% or include affiliations in footnotes:
\author[JILA]{Yan Zhou}

\author[JILA]{Kia Boon Ng}

\author[JH]{Lan Cheng}

\author[JILA]{Daniel N. Gresh\footnote{Current affiliation: Honeywell Quantum Solutions, 303 S. Technology Ct., Broomfield, CO 80021, USA}}

\author[MIT]{Robert W. Field}

\author[JILA]{Jun Ye}

\author[JILA]{Eric A. Cornell}

\address[JILA]{JILA, NIST and University of Colorado, and Department of Physics, University of Colorado, Boulder CO 80309, USA}
\address[JH]{Department of Chemistry, Johns Hopkins University, Baltimore, MD 21218, USA}
\address[MIT]{Department of Chemistry, Massachusetts Institute of Technology, Cambridge, MA 02139, USA}

\begin{abstract}
The molecular ion ThF$^+$ is the species to be used in the next generation of search for the electron's Electric Dipole Moment (eEDM) at JILA. The measurement requires creating molecular ions in the eEDM sensitive state, the rovibronic ground state $^3\Delta_1$, $v^+=0$, $J^+=1$. Survey spectroscopy of neutral ThF is required to identify an appropriate intermediate state for a Resonance Enhanced Multi-Photon Ionization (REMPI) scheme that will create ions in the required state. We perform broadband survey spectroscopy (from 13000 to 44000~cm$^{-1}$) of ThF using both Laser Induced Fluorescence (LIF) and $1+1'$ REMPI spectroscopy. We observe and assign 345 previously unreported vibronic bands of ThF. We demonstrate 30\% efficiency in the production of ThF$^+$ ions in the eEDM sensitive state using the $\Omega = 3/2$ [32.85] intermediate state. In addition, we propose a method to increase the aforementioned efficiency to $\sim$100\% by using vibrational autoionization via core-nonpenetrating Rydberg states, and discuss theoretical and experimental challenges. Finally, we also report 83 vibronic bands of an impurity species, ThO.

\end{abstract}

\begin{keyword}
    Thorium monofluoride, Thorium monoxide, electron EDM, LIF, REMPI, core-nonpenetrating Rydberg state, state-selective photoionization
\end{keyword}

\end{frontmatter}

%\linenumbers

\section{Introduction}

    The electron's electric dipole moment (eEDM) is a quantity with significant implications in the explanation of baryogenesis and dark matter, and also in fields such as particle physics and cosmology \cite{Ellis2016,Chupp2015,Stadnik2018}. The presence of a non-zero eEDM will give rise to anomalies in the Electron Spin Resonance spectra (ESR) in certain molecular levels. Currently the most sensitive searches \cite{cairncross2017precision,baron2014order,acme2018improved} for the eEDM are based on precision molecular spectroscopy. The JILA eEDM experiment is preparing to convert from using HfF$^+$ to using ThF$^+$. The most significant advantage of ThF$^+$ is that the eEDM sensitive state ($^3\Delta_1$, $v^+=0$, $J^+=1$) of ThF$^+$ is the electronic ground state \cite{gresh2016broadband,barker2012spectroscopic}, which means, in principle, that the ESR spectroscopy could exploit a very long coherence time. Furthermore, the effective electric field of ThF$^+$ (35.2~GV/cm \cite{Denis2015}) is 50\% larger than that of HfF$^+$ (24~GV/cm \cite{petrov2007theoretical,leanhardt2011high}), which promises a factor of 35.2/24=1.5 times increase of the eEDM sensitivity. Beyond measuring the eEDM, $^{229}$ThF$^+$ may also be a good candidate for determining the parity-forbidden Nuclear Magnetic Quadrupole Moment (NMQM), due to the large nuclear deformation of $^{229}$Th \cite{Flambaum2014a,Skripnikov2015}. The combination of eEDM and NMQM measurements from the same molecular system (different isotoplogues) would constrain new physics in both the hadronic and leptonic sectors of the standard model \cite{Sushkov1984}. Although all of the spectra presented in this paper are limited to $^{232}$ThF, it is straightforward to convert the measured molecular constants via the standard Born-Oppenheimer isotopologue scaling rule to those for $^{229}$ThF.
    
    In addition to enabling study of fundamental physics, spectroscopic studies of the high-lying excited states of ThF can also provide an interesting basis for comparison of experiment and theory. The molecular constants of highly excited states serve as comparison benchmarks for the development of relativistic quantum chemistry methods for molecules with actinide atoms \cite{Dolg2015}. Prior to the present work, experimental studies \cite{barker2012spectroscopic,heaven2014spectroscopy} of ThF were limited to REMPI spectroscopy within a small energy range below $T_e = 21500~\text{cm}^{-1}$.
    
    The first step of the JILA eEDM experiment is to prepare molecular ions in the eEDM sensitive state, which is $^3\Delta_1$, $v^+=0$, $J^+=1$ for both of our chosen molecular ions: HfF$^+$ and ThF$^+$. In the first generation of the JILA eEDM experiment, HfF$^+$ ions in the rovibronic ground state were created via vibrational autoionization of Rydberg states \cite{loh2011laser,cossel2012broadband,cairncross2017precision} by two-photon excitation, and then transferred to the eEDM sensitive state by Raman state transfer. In the current study, we take advantage of ThF$^+$ having $^3\Delta_1$ as its ground state, and demonstrate a more direct ion preparation method, without recourse to Raman state transfer. 
    
    To excite ThF from the ground state ($^2\Delta_{3/2}$) to the vibrationally autonizing Rydberg state ($>$51000~cm$^{-1}$), we implement a two-photon excitation scheme. The first photon ($>$32000~cm$^{-1}$) moves population to an intermediate state with excitation energy more than half of the ionization potential (IP), and the second photon is resonant with a vibrationally autoionizing Rydberg state. The key to this two-photon excitation scheme is to locate and identify an appropriate intermediate state. 
    
    The target intermediate state lies above half of the IP (25500~cm$^{-1}$) in the ultraviolet region. To optimize our setup for strong ThF signals, we check our system against previous ThF spectroscopic studies \cite{barker2012spectroscopic,heaven2014spectroscopy} in the visible region. As such, our spectroscopic survey covers the visible region (13000 to 16000~cm$^{-1}$ and 18000 to 20000~cm$^{-1}$) and the ultraviolet region (26000 to 44000~cm$^{-1}$). ThO appears as an impurity in our spectra, and we observe and fit multiple ThO vibronic bands as well.
    
    Among 345 identified ThF vibronic bands, we choose $\Omega$ = 3/2[32.85] as the intermediate state in the two-photon photoionization scheme. Combined with the convenient Nd:YAG 532 nm photon for the second transition, we create ThF$^{+}$ ions in a single electronic-vibrational state ($^3\Delta_1$, $v^{+}$=0), but a few rotational states ($J^{+}$=1-4). We also demonstrate that the final rotational distribution of ThF$^{+}$ ions is determined by the chosen rotational state of the intermediate state of neutral ThF molecules. By utilizing the lowest rotational state ($J$=3/2) of the intermediate state ($\Omega$ = 3/2[32.85]), we prepare 30\% of ThF$^+$ ions in the the eEDM sensitive state ($^3\Delta_1$, $v^+=0$, $J^+=1$). To increase this rotation-selectively population-efficiency toward $\sim$100\%, we propose to create ions by vibrational autoionization via core-nonpenetrating Rydberg states. Theoretical and experimental challenges are discussed in this paper.

\section{Experiment}

\begin{figure}[htb]
        \centering
        \includegraphics{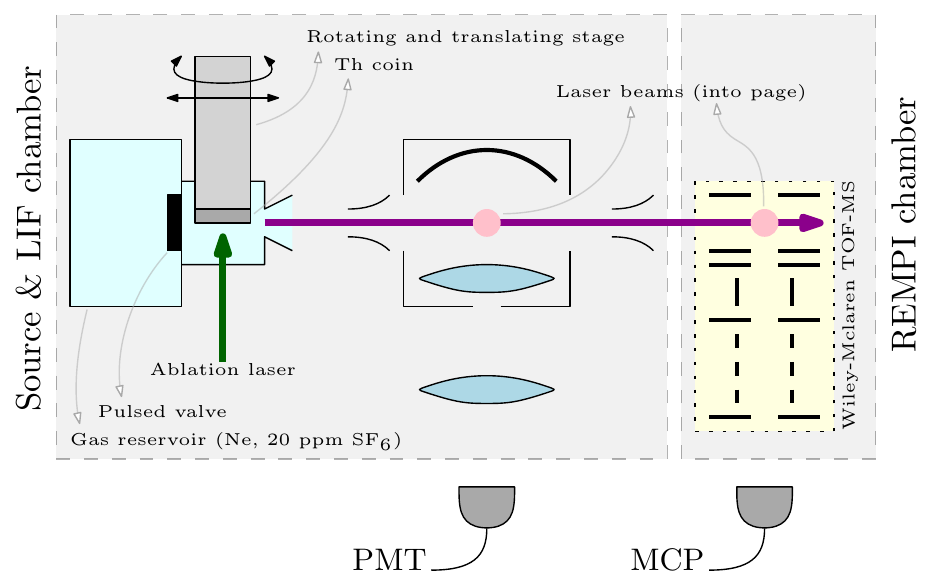}
        \caption{Schematic diagram of the experimental setup (not to scale). The molecular beam is generated in the same vacuum chamber in which the Laser-Induced Fluorescence (LIF) spectra are recorded. A detection cube consisting of a parabolic mirror and lenses is used to collect photons with a $\sim2\pi$ solid angle. The collected photons are sent to a photomultiplier. Resonant Enhanced Multi-Photon Ionization (REMPI) spectroscopy is performed  with a Time-of-Flight Mass Spectrometer (TOF-MS) located downstream in a separate chamber.}
        \label{fig:setupSchematic}
    \end{figure}

    A molecular beam of ThF is created by laser ablation of metallic thorium in a supersonic expansion of Neon seeded with SF$_6$. Laser-Induced Fluorescence (LIF) spectroscopy is performed in the same chamber as the molecular beam source, and Resonance Enhanced Multi-Photon Ionization (REMPI) spectroscopy is recorded in the second chamber, which is located downstream along the molecular beam axis, with a home-built time-of-flight mass spectrometer. A schematic diagram of the experimental setup is shown in Figure \ref{fig:setupSchematic}.
    The following sections describe the major components of our experimental setup in more detail.

    \subsection{Molecular beam source}
        Thorium plasma is generated by ablating thorium metal with a 5~ns, 2~mJ pulse of moderately focused 532~nm radiation ($\sim$150~$\upmu$m beam waist) from a Q-switched Nd:YAG laser. The thorium target is about 7~mm in diameter, and mounted on a slowly rotating and translating stage to provide continuously a fresh surface for ablation. The hot thorium plasma is chemically reacted with 20~ppm SF$_6$ in a neon buffer gas, which is kept at a stagnation pressure of 80~PSI. A home-built fast PZT valve releases a 50~$\upmu$s supersonic gas pulse into the vacuum chamber through a 3~mm expansion channel with diameter of 0.8~mm. The reacted ThF molecules in the beam are cooled to $\sim$10~K in both translational and rotational degrees of freedom via adiabatic expansion. We detect Th, ThF$_2$, ThF$^+$, ThO, and other species in the beam in addition to the desired ThF.
    
    \subsection{LIF experiment}
\begin{figure}[htb]
            \centering
            \begin{subfigure}[htb]{0.47\columnwidth}
                \includegraphics[width=\textwidth]{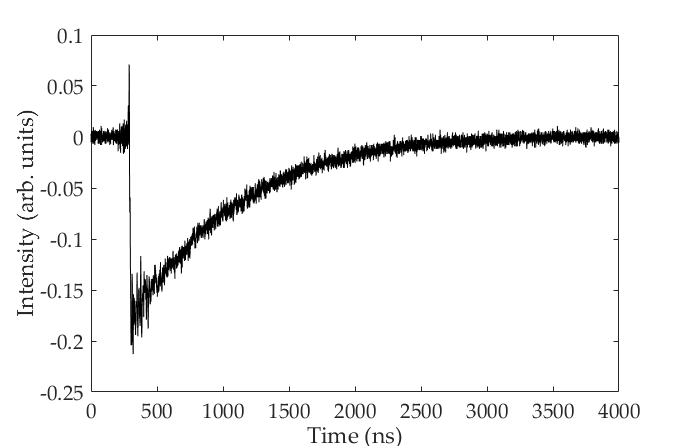}
                \caption{LIF}
                \label{fig:fluorescence}
            \end{subfigure}
            \vspace{5pt}
            \begin{subfigure}[htb]{0.47\columnwidth}
                \includegraphics[width=\textwidth]{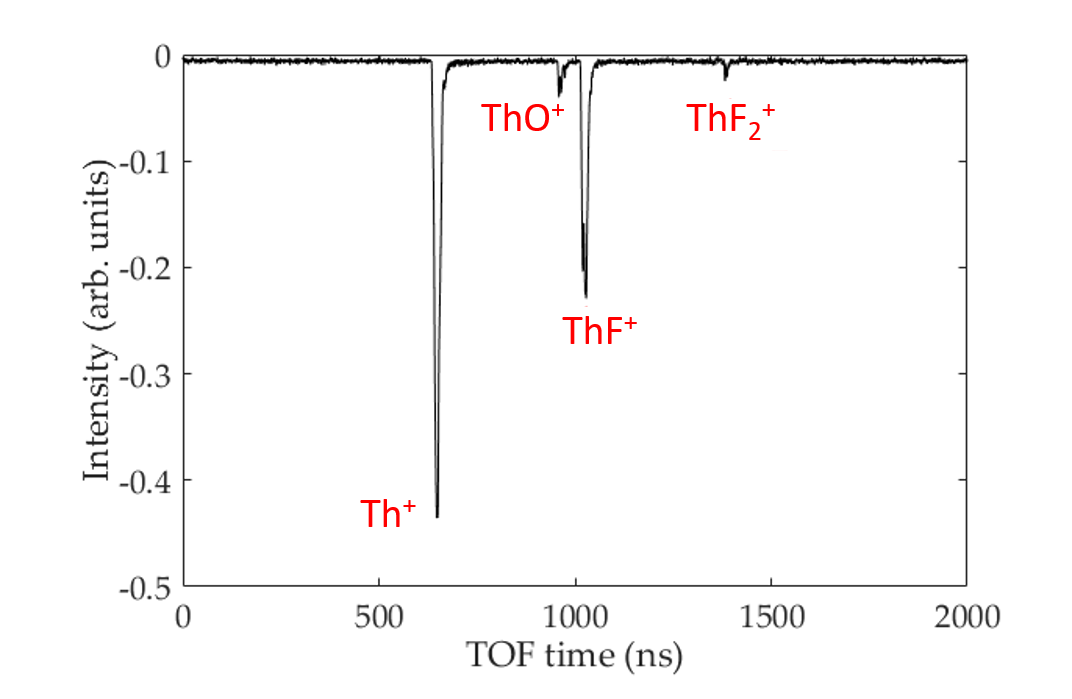}
                \caption{REMPI TOF-MS}
                \label{fig:rempi}
            \end{subfigure}
            \caption{Typical traces of LIF (a) and REMPI TOF-MS (b) experiments. From the LIF trace, we integrate the total intensity, and extract the fluorescence lifetime. The separation of species by mass in the REMPI TOF-MS trace allows us to distinguish the ion species to separate the ThF signals from the rest. Our mass resolution is sufficiently high to separate ThO$^+$ from ThF$^+$.}
            \label{fig:fluorescenceRempi}
        \end{figure}

        The molecular beam passes through a charged skimmer (50 volts applied) with a 3~mm aperture, which is 6~cm downstream from the molecular beam source, to extract only the coldest neutral molecules and deflect ions. A detection cube with $\sim$2$\pi$ collection solid angle is placed immediately after the skimmer. A broadband-coated (250 to 1200~nm) parabolic mirror above the beam is used to reflect fluorescence photons scattered upward into the photomultiplier tube (PMT) (Hamamatsu R3892) below. The excitation laser interrogates the molecular beam at the center of the detection cube. To reduce noise due to laser scattering, a fast switch gates the PMT on 10~ns after the excitation laser pulse to extract only the long-lived fluorescence signals ($>$50~ns). Figure \ref{fig:fluorescence} shows an example of the fluorescence trace, from which we integrate the total intensity, and extract the fluorescence lifetime.
    
    \subsection{REMPI TOF-MS experiment}
        In our LIF spectra, some ThF vibronic bands are contaminated by overlapping transitions of species like Th, ThO, and ThF$_2$. We use REMPI and a time-of-flight mass spectrometer (TOF-MS) to mass-isolate the ThF signals from the other molecular species. We scan the wavelength of the first photon of the REMPI process, while keeping the second photon fixed (532~nm or 355~nm). The REMPI spectra and LIF spectra in principle reveal the same information for ground to intermediate state transitions, but given the congested spectra, each method has its advantages: REMPI allows for rejecting lines from impurity species, while LIF allows for discrimination between congested transitions via fluorescence-lifetime record and better spectroscopic resolution. 
    
        In the REMPI TOF-MS experiment, the molecular beam passes through a skimmer with a 1~mm diameter aperture to enter a second vacuum chamber downstream with a $\sim3\times10^{-8}$~torr vacuum through differential pumping, as shown in Figure \ref{fig:setupSchematic}. This vacuum chamber houses a home-built TOF-MS in the orthogonal Wiley-Mclaren configuration. The molecular beam enters a region with a pair of parallel plates both charged at +1.5~kV, where the molecules are ionized by lasers propagating on an axis orthogonal to both the molecular beam and the TOF-MS axis. The molecules experience $<1$~V/cm stray electric field during the ionization process. Shortly after photoionization ($\approx$50~ns), the molecular ions are deflected by a 200~V/cm pulse toward a microchannel plate (MCP) detector, which is 75~cm downstream from the ionization region. The ion signal from the MCP is amplified by a transimpedance amplifier (Hamamatsu C9663), and recorded by a digital oscilloscope. Our TOF-MS has a fractional mass resolution of 1/500, which is sufficiently high to separate ThO$^+$ from ThF$^+$, as can be seen from a typical TOF-MS trace shown in Figure \ref{fig:rempi}.
    
    \subsection{Laser system and wavelength calibration}
        A tunable pulsed dye laser (Sirah Cobra-Stretch, 1800~grooves/mm single grating, 0.06~cm$^{-1}$ linewidth at 600~nm) is used to record a survey spectrum. The wavelength of the laser is monitored continuously by a wavemeter (Bristol 871B) during the scan. The wavemeter is referenced to a stabilized He-Ne laser with $\sim$50~MHz absolute accuracy. Some overlapping vibronic bands cannot be rotationally resolved by the pulsed dye laser, especially in the deep ultraviolet region. A narrow-band pulsed laser, which consists of a tunable cw-ring dye laser (Sirah, Matisse DR2) and a pulsed dye amplifier (Radiant Dye Amp, 150~MHz linewidth) is used to resolve these lines. The wavelength of the seeding laser is monitored by a high resolution wavemeter (High Finesse, WS7), which is referenced to an external-cavity diode laser locked to a $^{87}$Rb transition at 384.227982~THz.
        
\section{Results \& Analysis}
    
    \subsection{Data set}
        We record survey scans over the ranges of 13000 to 16000~cm$^{-1}$, 18000 to 20000~cm$^{-1}$, and 26000 to 44000~cm$^{-1}$ using multiple laser dyes (LDS 698, DCM, Coumarin 540A, Coumarin 503, Coumarin 480 and Coumarin 460), and as necessary, frequency-doubled with a BBO crystal. 345 previously unreported vibronic bands of ThF are recorded. The fitted rotational constants are presented in Tables \ref{tab:all1} to \ref{tab:all14}.

\begin{figure}[htb]
            \centering
            \begin{subfigure}[htb]{0.47\columnwidth}
                \includegraphics[width=\textwidth]{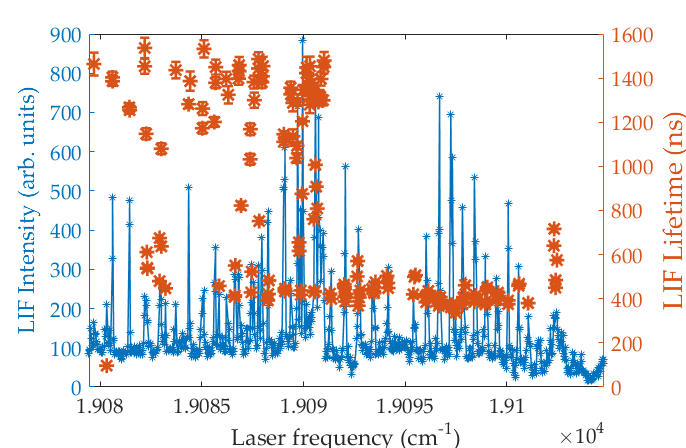}
                \caption{Decoupling by fluorescence lifetime}
                \label{fig:lifetimes}
            \end{subfigure}
            \vspace{5pt}
            \begin{subfigure}[htb]{0.47\columnwidth}
                \includegraphics[width=\textwidth]{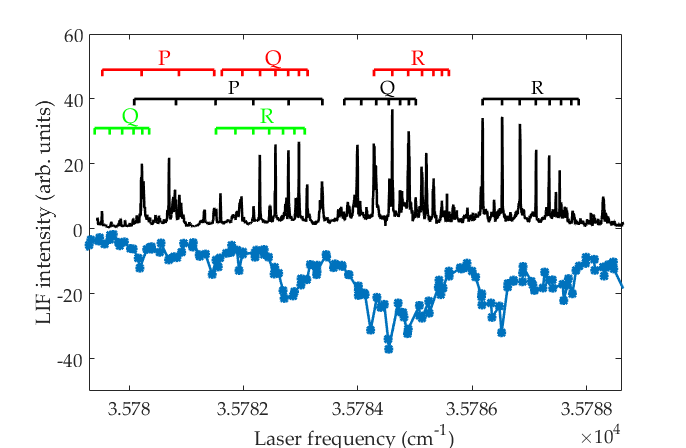}
                \caption{High resolution scan}
                \label{fig:highResolution}
            \end{subfigure}
            \caption{Two methods to resolve congested regions of the spectrum. (a) Radiative lifetimes permit separation of two overlapping bands. Lines from the same vibronic band should all have the same fluorescence lifetime. In this plot, transitions are from two vibronic bands with lifetimes of 400 ns and 1400 ns, respectively. (b) High resolution scan with a narrow-band pulsed laser. Top and bottom traces correspond to scans performed respectively by the narrow-band pulsed laser (0.005~cm$^{-1}$) and normal pulsed laser (0.1~cm$^{-1}$).}
            \label{fig:congested}
        \end{figure}

        The density of electronic states of ThF is much higher than that of HfF. The electronic states below the ionization threshold in ThF can be formed nominally by distributing the valence electrons among the $7s$, $6d$, $7p$, and $5f$ orbitals. In contrast, the electronic states in HfF are formed by distributing the valence electrons among $6s$, $6p$, and $5d$ orbitals only, as the $4f$ shell of Hf is fully occupied and thus is not accessible. Furthermore, the $7s$ shell in ThF is much more polarizable than the $6s$ shell in HfF, as the latter is significantly stabilized by the lanthanide contraction. In this experimental work, there is about one vibronic band every 40~cm$^{-1}$ on average. It is not uncommon to find two or more ThF vibronic bands overlapping with each other. We use two methods to disentangle the overlapping bands: \begin{inparaenum}[(i)] \item differentiating rotational bands with respect to their different fluorescence lifetimes, as shown in Figure \ref{fig:lifetimes}; and \item high resolution spectroscopy ($<$0.005~cm$^{-1}$ resolution) with a narrow-band injection-seeded pulsed laser, as shown in Figure \ref{fig:highResolution}. \end{inparaenum}
        
        In the visible region (13000 to 20000~cm$^{-1}$), all vibronic bands are well resolved, and we identify vibrational progressions using vibrational combination difference analysis. In the ultraviolet region (26000 to 38000~cm$^{-1}$), most vibronic bands can be resolved, but we are only able to identify a few vibrational progressions. In the deep ultraviolet region (38000 to 44000~cm$^{-1}$), only a few vibronic bands have been resolved, and no vibrational progression can be identified. 
        
        Although we do not intentionally introduce oxygen containing compounds to our molecule production region, we see significant numbers of strong ThO bands. Both because of their intrinsic interest, and also the need to identify lines which do not belong to ThF bands, we have fitted 83 ThO bands. Previous experimental works \cite{ThOgatterer1957molecular,ThOvon1970rotational,ThOwentink1972isoelectronic,ThOedvinsson1965g,ThOzare1973direct,ThOackermann1973high,ThOhildenbrand1974mass,NISTdatabase} have detected bands and made electronic level assignments in ThO with energies $T_e$ as high as 22683~cm$^{-1}$. We see many previously detected vibronic levels and also extend the ThO survey to vibronic levels as high as $T_e=41,000~\text{cm}^{-1}$. These bands are presented in Tables \ref{tab:allThO1} to \ref{tab:allThO4}.

    \subsection{Fitting models}
        
        Rotational bands were contour-fitted to the form:
        \[ \nu = \nu_0 + F'(J') - F''(J''), \]
        where $\nu$ is the measured frequency, $\nu_0$ is the vibronic band origin, and $F'(J')$ and $F''(J'')$ are the rotational energies of the upper and lower states, respectively, combined with H\"onl-London factors, which are described as in equation (\ref{eq:HLc}).
        
        Signal-to-noise is in general insufficient to allow a stable fit to a single band with simultaneous variation of both upper and lower rotational constants. With this limitation, and in view of the fact that for all the ThO bands, the lower state is consistent with being the $\Omega''=0$ ground electronic state, we used previously measured values of the rotational constant, $B''$, in the relation:
        \[ F''(J'') = B'' J'' (J''+1), \]
        where $J''$ is the rotational quantum number of the ground state. A similar situation applies for ThF, except that the ground state is $\Omega''=3/2$, and the rotational constants are determined by a global fit to many bands, as discussed in Section \ref{sec:family}. For bands that have not been assigned to a vibrational progression so that we cannot be certain of the lower state vibrational quantum number, $v''$, we consider that the large majority of the lower-state population is in either $v''=0$ or $v''=1$, and, as a compromise, fix $B''$ to be the average of the $B''$ for those two vibrational levels. For low rotational temperatures, there was no need to fit the lower states with the centrifugal rotational term, $D$, or with non-vanishing $\Omega$-doubling.
        
        As for the upper states, we fit to three main classes: \begin{inparaenum}[(i)] \item transitions that are well described by Hund's case (c), \item transitions with $^2\Pi$ character, and \item transitions described by Hund's case (b). \end{inparaenum} The fitting routine for each of these classes are described in the following sections.
        
        \subsubsection{Hund's case (c)}
            We expect most of the electronic states in ThF and ThO to be subject to a spin-orbit interactions much larger than the rotational spacing, due to the presence of the heavy thorium atom. This hierarchy in interaction energies holds especially in the low-$J$ region. As such, these states are well described by Hund's case (c).
            
            We perform a fit of the vibronic bands with the upper state described by the following Hamiltonian:
            \begin{equation}\label{eq:caseC}
                F'(J') = B' J' (J'+1),
            \end{equation} 
            where $B'$ and $J'$ are the rotational constant for the excited state of ThF, and the rotational quantum number of the excited electronic states, respectively. The centrifugal distortion rotational constant, $D'$, cannot be fitted to our rotational bands, because \begin{inparaenum}[(i)] \item our beam has a $\sim$10~K rotational temperature, which is not high enough to populate the high $J''$'s for a satisfactory fit to $D'$, and \item we have relatively low spectroscopic resolution. \end{inparaenum} On a similar note, the $\Omega$-doubling for the low-$J'$ lines is much smaller than our spectroscopic resolution, hence we ignore it in our fitting model.
            
\begin{figure}[htb]
                \centering
                \begin{subfigure}[htb]{0.47\columnwidth}
                    \includegraphics[width=\textwidth]{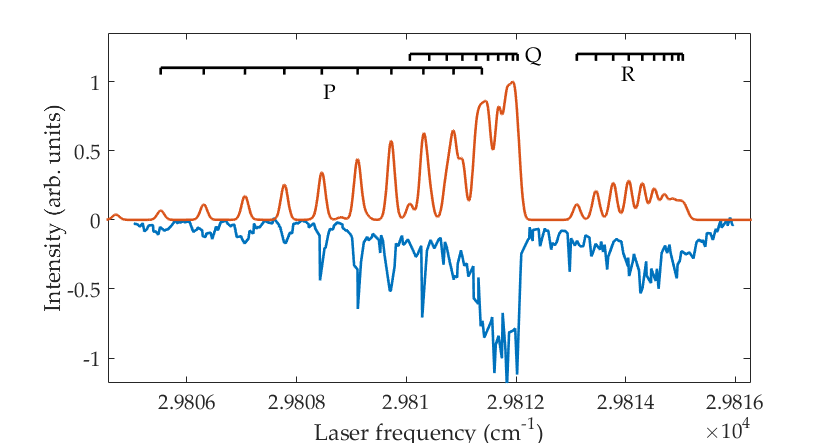}
                    \caption{$\Omega'=1/2$}
                \end{subfigure}
                \vspace{5pt}
                \begin{subfigure}[htb]{0.47\columnwidth}
                  \includegraphics[width=\textwidth]{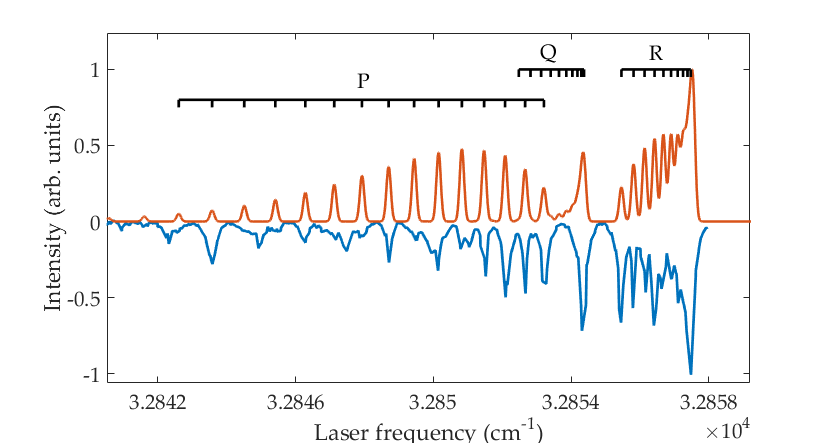}
                    \caption{$\Omega'=3/2$}
                \end{subfigure}\\
                \begin{subfigure}[htb]{0.47\columnwidth}
                    \includegraphics[width=\textwidth]{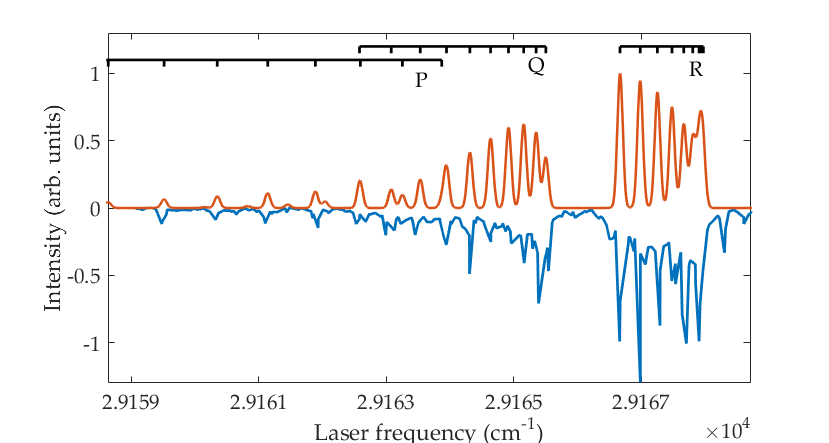}
                    \caption{$\Omega'=5/2$}
                \end{subfigure}
                \caption{Plots of typical bands with various values of $\Omega'$. The values of $\Omega'$ for our bands can be determined by observation of the lowest-$J'$ lines in the $P$ branch, and the relative intensities of the $PQR$ branches. The trace in red (blue) corresponds to results from fitting (data).}
                \label{fig:Omega}
            \end{figure}

            We assign $\Omega'$, the projection of the total angular momentum onto the inter-nuclear axis, to the transitions by referring to the low $J'$ lines in the $P$ branch and the relative intensities of the $PQR$ branches. The relative intensities of the $PQR$ branches are given by the H\"onl-London factors ($\text{HL}_{\Omega' J', \Omega'', J''}$) shown in the following equation \cite{WATSON20085}:
            \begin{align}\label{eq:HLc}
                \text{HL}_{\Omega' J', \Omega'', J''} = &\left( 1 + \delta_{\Omega' 0} + \delta_{\Omega'' 0} - 2 \delta_{\Omega' 0} \delta_{\Omega'' 0} \right) \left( 2 J' + 1 \right) \left( 2 J'' + 1 \right) \notag\\
                &\times \begin{psmallmatrix} J' & 1 & J'' \\ -\Omega' & (\Omega' - \Omega'') & \Omega'' \end{psmallmatrix}^2,
            \end{align}
            where symbols with double and single primes correspond to the lower and upper states, respectively; $\delta_{\Omega 0}$ is the Kronecker delta factor, and the last term is the Wigner 3$j$ symbol. Plots of typical rotational bands with various values of $\Omega'$ are shown in Figure \ref{fig:Omega}.
        
        \subsubsection[Doublet-Pi states]{$^2\Pi$ states}
\begin{figure}[htb]
                \centering
                \includegraphics[width=0.75\columnwidth]{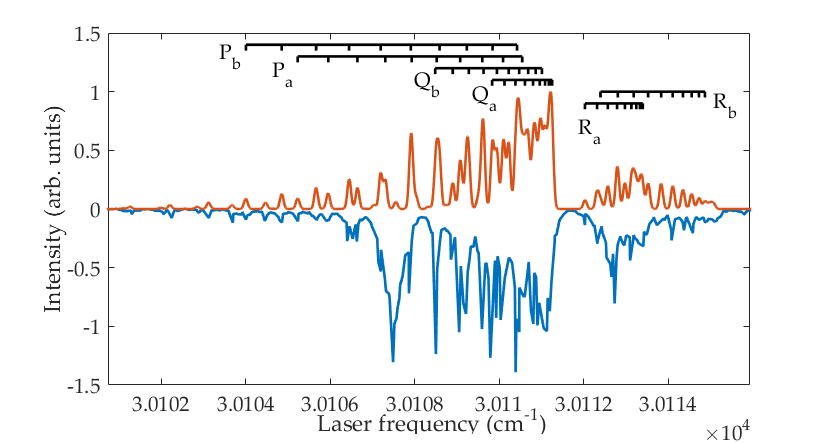}
                \caption{A ThF band with resolvable $\Lambda$-doubling.}
                \label{fig:Lambda}
            \end{figure}

            The next class of observed transitions requires us to include a dominant $\Lambda$-doubling term in our fitting model, which implies a strong $^2\Pi$ character. Since we do not have high enough spectroscopic resolution to resolve the e/f symmetry components of the lower state of the transition, we cannot determine the absolute parities of the upper state of the transition. Following the convention used in Loh et al.\ \cite{loh2012rempi}, we label related $\Lambda$-doubling pairs as $a/b$ instead of $e/f$. The upper state can be modeled by the following Hamiltonian:
            \begin{equation}\label{eq:doubletPi}
                F'_{a/b}(J') = B' J' (J'+1) \mp \frac{1}{2} (p+2q) \left( J' + \frac{1}{2} \right),
            \end{equation}
            where $(p+2q)/2$ is the $\Lambda$-doubling constant. The upper (lower) sign corresponds to the $a$ ($b$) symmetry component. An example of a band with resolvable $\Lambda$-doubling is shown in Figure \ref{fig:Lambda}.

        \subsubsection{Hund's case (b)}
            The third class of transitions that we observe consist of two sets of $PQR$ branches, which is characteristic of states with Hund's case (b) character, where the absence of a zeroth-order spin-orbit splitting results in the uncoupling of the spin from the rotational angular momentum. We perform the fit to the transition with the upper state described by the following model:
            \begin{align*}
                F'_a(N') &= B' N' (N'+1) - \frac{\gamma}{2}(N+1), \\
                F'_b(N') &= B' N' (N'+1) - \frac{\gamma}{2}N,
            \end{align*}
            where $N'$ is the rotational quantum number excluding spin, and $\gamma$ is the spin-rotation constant. Since the pattern-forming rotational quantum numbers for the lower and upper states are $J''$ and $N'$, respectively, angular momentum selection rules allow for two sets of $PQR$ branches:
            \begin{align*}
                P_{a/b} &: \left(N' \mp \frac{1}{2} \right) - J'' = -1, \\
                Q_{a/b} &: \left(N' \mp \frac{1}{2} \right) - J'' = 0, \\
                R_{a/b} &: \left(N' \mp \frac{1}{2} \right) - J'' = 1,
            \end{align*}
            where the upper (lower) sign corresponds to the $a$ ($b$) state. 
            
\begin{figure}[htb]
                \centering
                \includegraphics[width=0.75\columnwidth]{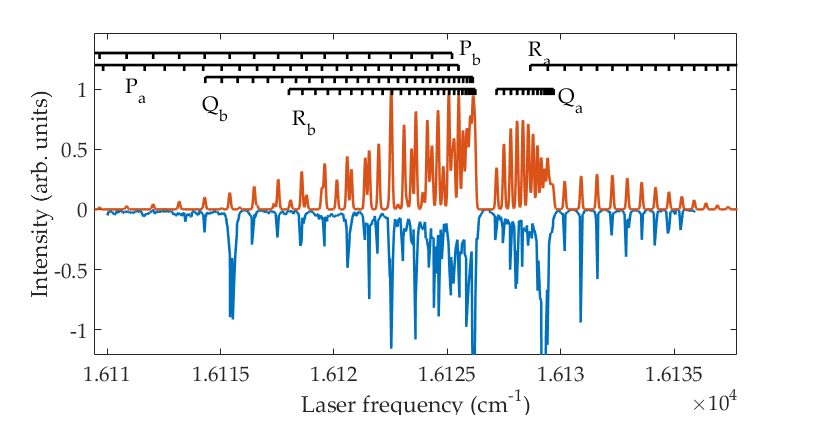}
                \caption{A typical ThF band fitted to Hund's case (b).}
                \label{fig:CaseB}
            \end{figure}

            For a contour fit, the H\"onl-London factor for Hund's case (c) to case (b) transitions used for the fitting is as follows:
            \[ \text{HL}_{\Lambda'J'N',\Omega''J''} = \left| \sum_{\Sigma', \Omega'} \sqrt{2N'+1} \, (-1)^{J'+\Omega'} \sqrt{\text{HL}_{\Omega' J', \Omega'' J''}} \begin{psmallmatrix} S & N' & J' \\ \Sigma' & \Lambda' & -\Omega' \end{psmallmatrix} \right|^2, \]
            where $\text{HL}_{\Omega' J', \Omega'' J''}$ is the H\"onl-London factor for a Hund's case (c) to case (c) transition, as shown in equation \eqref{eq:HLc}; the last term is the Wigner 3$j$ symbol; $\Lambda'$ is the projection of the orbital angular momentum onto the internuclear axis; and $S$ is the total spin of the electrons, which is 1/2 for all of the transitions that we have fitted. A typical band fitted with Hunds's case (b) is shown in Figure \ref{fig:CaseB}.
        
    \subsection{Vibrational progression}\label{sec:family}
        Using a combination difference analysis, we tentatively group vibronic bands that share the same upper and lower electronic states, but different vibrational levels, into sets of vibrational progressions. The combination differences of the lower states are known from previous work \cite{barker2012spectroscopic,heaven2014spectroscopy}. The criteria for extracting combination differences for the upper states are: \begin{inparaenum}[(i)] \item vibronic bands that belong to the same electronic state should have similar rotational constants; and \item vibrational constants should agree with \emph{ab initio} calculations within a reasonable range ($500<\omega_e<700~\text{cm}^{-1}$, $0<\omega_{eXe}<10~\text{cm}^{-1}$, similar radiative lifetime). \end{inparaenum} 
        
        For transitions that we associate with a vibrational progression, we redo the fit to the vibronic band using the rotational constant ($B_{v''}$) for the relevant vibrational quantum number ($v''$) of the lower state in the transition. From these fits, we extract the molecular constants $\nu_e$, $B_e$, $\alpha_e$, $\omega_e$, and $\omega_{eXe}$. These constants are related to the fitting constants ($\nu_0$, $B_v$, and $v$, which are defined in previous sections) as such:
        \begin{subequations}\label{eq:vibrationalParameters}
            \begin{align}
                \begin{split}\nu_0 &= \nu_e + \omega'_e \left(v'+\frac{1}{2}\right) - \omega'_{eXe}\left(v'+\frac{1}{2}\right)^2 \\
                 &\hphantom{= \nu_e} - \omega''_e \left(v''+\frac{1}{2}\right) + \omega''_{eXe}\left(v''+\frac{1}{2}\right)^2 , \end{split}\\
                B_v &= B_e - \alpha_e\left(v+\frac{1}{2}\right).
            \end{align}
        \end{subequations}
        
        A comprehensive list of the upper state vibrational progressions that we have tentatively assigned is shown in Table \ref{tab:families}, with lower state values in Table \ref{tab:groundState}.
        \begin{table}[htb]
            \centering
            \begin{tabular}{| c | c c c c c c |}
                \hline
                & $\nu_e$ & $B_e$ & $\alpha_e$ & $\omega_e$ & $\omega_{eXe}$ & $\frac{p+2q}{2}$ \\
                \hline
                 $\Omega=3/2$ [14.09] & 14093.80(3) & 0.2276(5) & - & 557.25(2) & - & - \\
                 $\Omega=3/2$ [15.18] & 15180.19(1) & 0.2221(9) & - & 584.00(4) & - & - \\
                 $\Omega=3/2$ [18.62] & 18616.19(2) & 0.2179(5) & 0.0013(2) & 517.54(3) & 2.00(1) & - \\
                 $\Omega=3/2$ [19.98] & 19978.19(3) & 0.2172(4) & 0.0007(3) & 533.05(3) & - & - \\
                 $\Omega=3/2$ [20.40] & 20397.40(2) & 0.2126(3) & - & 569.47(2) & - & - \\
                 $^2\Pi_{1/2}$ [27.98] & 27977.94(4) & 0.2173(8) & - & 563.13(4) & - & 0.009(6) \\
                 $\Omega=3/2$ [28.37] & 28370.52(3) & 0.2138(5) & - & 577.59(3) & - & - \\
                 $^2\Pi_{1/2}$ [29.22] & 29223.19(6) & 0.2175(3) & - & 580.5(1) & 3.80(8) & 0.022(3) \\
                 \hline
            \end{tabular}
            \caption{Comprehensive list of tentative assignments of vibrational progressions. Vibrational progressions are groups of transitions which share the same upper and lower electronic states, but different vibrational levels. The molecular constants are defined in equations \eqref{eq:doubletPi} and \eqref{eq:vibrationalParameters}. Units for all the above constants are quoted in cm$^{-1}$. 90\% confidence intervals, which come from a convolution of spectroscopic fitting uncertainties and vibrational progression fitting uncertainties, are quoted in parentheses. Large fitting uncertainties for the rest of the vibrational progressions prevent us from extracting their corresponding $\alpha_e$, so $\alpha_e$ is set to zero for our fitting purposes.}
            \label{tab:families}
        \end{table}
        For progressions with only two different $v'$, e.g. $v'=0$ and $v'=1$, we conventionally set $\omega_{eXe}$ to zero so that $\omega_e$ is just the splitting between the relevant rotational band origins of the upper state. 90\% confidence intervals, which come from a convolution of fitting uncertainties and vibrational progression fitting uncertainties, are quoted in parentheses in Table \ref{tab:families}.
        
        Since the bands in our data set have relatively low resolution and high state density, there is a risk that transitions are incorrectly grouped, and that apparent progressions occur only due to coincidence. Hence, more experiments are required to confirm these preliminary assignments, and we leave this as a challenge to our fellow spectroscopists.
        
        The aforementioned analysis allows us to group most of the vibronic bands found in the visible region into vibrational progressions. In the ultraviolet region, however, we are unable to identify any definite vibrational progressions. Missing transitions in vibrational progressions could be explained by perturbations of these high energy levels, which are discussed in Section \ref{sec:perturbationDeepUV}.

    \subsection{The Ground State of ThF}
        All of the ThF transitions that we have assigned seem to have the $^2\Delta_{3/2}$ ground electronic state as the lower level. The first excited state of ThF is $^2\Delta_{5/2}$, which is $2500~cm^{-1}$ higher in energy and has not been observed in our experiments\cite{barker2012spectroscopic}. As such, we use all of the identified transitions (shown in Table \ref{tab:families}) to perform a global fitting to extract the molecular constants of the ground state precisely. 
        \begin{table}[htb]
            \centering
            \begin{tabular}{| c | c c c |}
                \hline
                & This work (exp.) & Barker et al.\ \cite{barker2012spectroscopic} & This work (theory) \\
                \hline
                 $\omega_e$ (cm$^{-1}$) & 601.00(2) & 605(15) & 598.8 \\
                 $\omega_{eXe}$ (cm$^{-1}$) & 2.07(3) & - & 2.1 \\
                 $B_e$ (cm$^{-1}$) & 0.2339(2) & 0.237(5) & 0.2325 \\
                 $\alpha_e$ (cm$^{-1}$) & 0.0014(3) & - & - \\
                 $r_e$ (\AA) & 2.026(3) & 2.01(3) & 2.032 \\
                 \hline
            \end{tabular}
            \caption{Comparison of molecular constants of the $^2\Delta_{3/2}$ ground state of ThF in experiment and theory. 90\% confidence intervals are quoted in parentheses. The parameter $r_e$ is the equilibrium internuclear distance, and the other molecular constants are defined in equation \eqref{eq:vibrationalParameters}.}
            \label{tab:groundState}
        \end{table}
        
        In addition, we have performed \emph{ab initio} calculations for the molecular constants using the CFOUR program \cite{CFOUR}. A spin-orbit (SO) version \cite{X2CSOCC2018} of the coupled-cluster singles and doubles with a non-iterative treatment of triple excitations [CCSD(T)] method \cite{Pople1989} has been adopted using the exact two-component (X2C) Hamiltonian \cite{Dyall1997, Saue2007} with atomic mean-field SO integrals \cite{Liu2018} and the uncontracted ANO-RCC basis sets \cite{Faegri2001,ROOS2005}. The $5s$, $5p$, $5d$, $6s$, $6p$, and $7s$ electrons of Th as well as $2s$ and $2p$ electrons of F have been correlated together with virtual spinors with energies below 1000 Hartree. 
        %The present study has taken into account spin-orbit effects as well as the correlation of the sub-valence electrons, and thus is more rigorous than previous calculations in the literature \cite{Lester2013}. 
        These calculations with the correlation of semi-core electrons and extensive virtual space (37 electrons and 679 virtual spinors) have been expedited using the recently developed semi-atomic-orbital based algorithm for SO-CCSD(T) \cite{X2CSOCC2018}. The local potential energy curve has been fit to an eighth-order polynomial to obtain linear through quartic force constants, which determine the corresponding parameters in a Morse potential as well as the molecular constants. The X2CAMF-CCSD(T) results for the ground state molecular constants of ThF are summarized in the fifth column of Table \ref{tab:GSCOMP}. To demonstrate the spin-orbit effects for these parameters, we have also performed scalar relativistic CCSD(T) calculations based on the spin-free exact two-component theory in its one-electron variant (SFX2C-1e) \cite{SFX2C1e,Cheng2011} using the ANO-RCC-unc basis sets. In addition, the SFX2C-1e-CCSD(T) results with the correlation of only the Th 6s, 6p, 7s, 6d electrons and the F 2s and 2p electrons are also presented in Table \ref{tab:GSCOMP} (the "SFX2C-1e/LC" results) to be directly compared with the effective-core-potential  (ECP)  calculations in Ref. \cite{Lester2013}. 
        
        \begin{table}[htb]
            \centering
            \begin{tabular}{| c | c c c c |}
                \hline
               % & ECP/aVTZ         & Barker et al.\ \cite{barker2012spectroscopic} & This work (theory) \\
		&  ECP/LC [37] %\cite{Lester2013}      
		& SFX2C-1e/LC  & SFX2C-1e/SC  & X2CAMF/SC    \\
                \hline
                 $\omega_e$ (cm$^{-1}$)     & 598.8     & 602.4    & 605.3    & 598.8     \\
                 $\omega_{eXe}$ (cm$^{-1}$) & 2.1       & 2.1      & 2.1      & 2.1       \\
                 $r_e$ (\AA)                & 2.032     & 2.034    & 2.030    & 2.032     \\
                 \hline
            \end{tabular}
            \caption{The CCSD(T) results for the molecular constants of the $^2\Delta_{3/2}$ ground state of ThF.
	    The parameter $r_e$ is the equilibrium internuclear distance, and the other molecular constants 
	    are defined in equation \eqref{eq:vibrationalParameters}.
    ``LC'' and ``SC'' represent the correlation of 19 and 37 electrons in the CCSD(T) calculations, respectively.
    The effective-core-potential (ECP) calculations in Ref. [37] %\cite{Lester2013} 
    have used the aug-cc-pVTZ basis sets,
    while the SFX2C-1e and X2CAMF calculations of this work have adopted the ANO-RCC-unc basis sets.}
            \label{tab:GSCOMP}
\end{table}

        It can be seen that the SO contributions (the difference between X2CAMF and SFX2C-1e results) amount to -6.5 cm$^{-1}$ for the harmonic frequency and -0.002 {\AA} for the equilibrium bond length. The contributions from the correlation of sub-valence electrons (the difference between SFX2C-1e/SC and SFX2C-1e/LC results) and the corrections for ECP (the difference between SFX2C-1e/LC and ECP/LC results) are of similar magnitude.  Interestingly, a fortuitous cancellation of these three types of contributions is observed for the molecular parameters studied here and this leads to the close agreement of the present X2CAMF-CCSD(T) results and the ECP-CCSD(T) results in Ref. \cite{Lester2013}. 
        
        %Note that this is equivalent to the use of these force constants in a vibrational %second-order perturbation theory calculations for the solution of the nuclear %Schr{\"o}dinger equation. 
        The X2CAMF-CCSD(T) results of this work are presented in Table \ref{tab:groundState} together with the experimental results as well as previous measurements. Agreement between measured and computed values has been observed for all parameters here. The remaining errors for the bond length and the harmonic frequency are around 0.005 {\AA} and several cm$^{-1}$ (less than 1\% of the total value), which are consistent with the typical errors of the CCSD(T) method in calculations for organic molecules. This demonstrates that the accurate treatment of relativistic effects provided by the X2C method can extend the accuracy and capability of CCSD(T) to a heavy-metal containing system.
    
    \subsection{Chosen transition for two-photon ionization scheme}
        We choose the $\Omega=3/2$ [32.85] state as the intermediate state for the creation of ThF$^+$ in the $^3\Delta_1$, $v^+=0$ state through a resonance-enhanced two-photon ionization scheme. This choice is motivated by several considerations. 
        
        First, the photoionization scheme involves two photons of different colors, of which the second photon must be more intense than the first one in order to saturate the much weaker transition. Since this chosen intermediate state lies above IP/2, non-resonant ionization from ground state of ThF by the more intense second photon is greatly suppressed as such excitation must be a three-photon process.
        
        Secondly, the transition from the ground state to the $\Omega=3/2$ [32.85] state is a relatively strong transition in a wavelength region that is easily accessible. 
        
        Finally, a convenient 532~nm photon (from the second harmonic of Nd:YAG laser) can be used as the second photon, which excites into a broad autoionizing resonance that lies only 62~cm$^{-1}$ above the ionization potential for excitation into the $^3\Delta_1$, $v^+=0$ state in ThF$^+$. Since the lowest energy vibronic excited state ($^1\Sigma^+$, $v^+=0$) lies 314~cm$^{-1}$ above the ground state, energy considerations restrict all ions that all ions formed must be in the ground vibronic state, which is the eEDM sensitive state. 
        
\begin{figure}[htb]
            \centering
            \includegraphics[width=0.9\columnwidth]{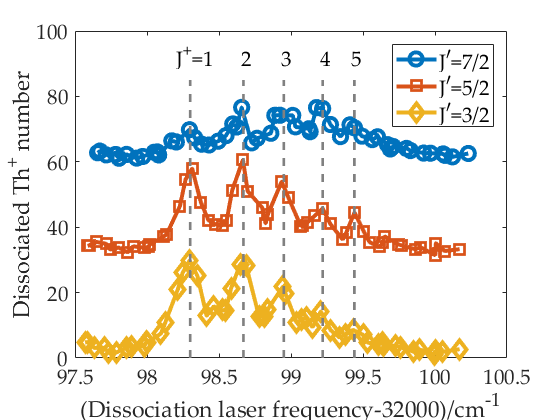}
            \caption{$J^+$-distributions of ThF$^+$ ions created from different $J'$ intermediate states of ThF. The rotational distribution of the ThF$^+$ ions shifts toward higher $J^+$ states as the $J'$ value of the intermediate state is increased. For easy comparison purposes, the red and blue traces are offset upward from zero by 30 and 60, respectively. These traces are scans of the $R$ branch of the dissociated intermediate state used in the resonance-enhanced multi-photon dissociation spectrum, which will be described in our upcoming paper \cite{dissociation}.}
            \label{fig:Jdependence}
        \end{figure}

        We verify the formation of ions in the ThF$^+$ground vibronic state by state-selective resonantly enhanced multi-photon dissociation, which will be described in our upcoming paper \cite{dissociation}. Using the dissociation readout technique, we also show that the distribution of the ion population across the different $J^+$ levels depends on the $J'$ quantum number of the intermediate state used in the REMPI process. The $J^+$-distributions of the ions created with different $J'$ intermediate states are shown in Figure \ref{fig:Jdependence}.
        
        Figure \ref{fig:Jdependence} shows that the rotational distribution of the ions shifts toward higher $J^+$ states when a higher $J'$ intermediate state is used. Since the electron that is lost during the ionization process carries a fixed range of angular momenta, we expect to see a higher final angular momentum distribution for the ions when an intermediate state of higher angular momentum is used, and \emph{vice versa} \cite{Xie1990a}. Since the eEDM sensitive state is the lowest rotational state ($J^+=1$), we use the $J'=3/2$ state as our intermediate state for the REMPI process, which results in 30\% of the ions being produced in the desired $J^+=1$ state.

\section{Further Discussion}
    \subsection{Prospects of rotationally state-selective photoionization}
        Inasmuch as the aforementioned two-photon photoionization scheme creates ThF$^+$ ions in the ground vibronic state, the ion population is spread across several rotational levels (typically $J^+=1-4$). However, only the $J^+=1$ state is useful for the eEDM measurement. All of the ions in the other rotational states do not contribute to the eEDM sensitivity, but are nonetheless co-trapped. These background ions in the trap contribute to decoherence of the eEDM measurement by ion-ion collisions with the useful ions, hence reducing the eEDM measurement sensitivity.
        
        To improve rotational state purity in the photoionization process, we propose a scheme to achieve rotationally-selective photoionization via a core-nonpenetrating Rydberg state, where the orbital angular momentum of the Rydberg electron will be $l\geq4$. In such a scheme, we would prepare the neutral molecule in a core-nonpenetrating long-lived ($\tau>$10 ns) vibrationally-autoionizing Rydberg state that belongs to a Rydberg series that converges to the ion's $v^+=1$ state, with high orbital angular momentum ($l\geq4$) by a resonant two-photon excitation. The Rydberg molecule is ionized by the transfer of energy from the vibration of the ion-core to the Rydberg electron. There will be no hard collision between the Rydberg electron and the ion core, because the Rydberg electron is always kept far away from the ion-core by the $l(l+1)/2\mu r^2$ centrifugal barrier, where $\mu$ is reduced mass of the Rydberg electron on ThF$^{+}$, and r is distance between the Rydberg electron and ThF$^+$ ion core. Hence the energy transfer mechanism will be predominantly a long-range electric dipole or quadrupole interaction, which preserves the ion-core rotational state during the autoionization process. Thus, the final rotational distribution of ThF$^+$ would be the same as the rotational distribution of the neutral ThF Rydberg state, which can be selectively populated by the optical-optical double resonance method.
        
        Since the $^1\Sigma^+$ ($s^2$) ion-core is more compact than the $^3\Delta_1$ ($sd$) ion-core, and gives rise to a simpler $s^2(^1\Sigma^+)nl$ electronic structure than $sd(^3\Delta_1)nl$, we propose to prepare ThF$^+$ in the $^1\Sigma^+$ state instead of $^3\Delta_1$ in the first step of this scheme. We can then apply Raman-type transitions to move the population from $^1\Sigma^+$ to the eEDM sensitive $^3\Delta_1$ state. The Raman transfer scheme has already been demonstrated in the HfF$^+$ experiment \cite{cossel2012broadband,cairncross2017precision}.
        
        In practice, implementing this scheme is more challenging than we expected. Since the ThF$^+$ $^1\Sigma^+$ state lies above that of $^3\Delta_1$, which has a higher degeneracy, the probability of excitation to a Rydberg state that converges to $^3\Delta_1$ is much greater than one that converges to $^1\Sigma^+$. The use of an appropriate intermediate state in this two-step excitation scheme is critical to increasing the probability of the preparation of Rydberg states that converge to the $^1\Sigma^+$ state. Since the ground electronic structure of ThF is $s^2d$, an intermediate state with $s^2f$ character is preferable to $spd$ in providing access to Rydberg states with the $^1\Sigma^+$ $s^2$ ion-core. Identifying such a state with a specific electronic configuration requires more theoretical work, which is described in the next section.
    
    \subsection{Ab initio calculations}
        We have been unable to find any prior work on \emph{ab initio} calculations of high-lying excited states of actinide-containing molecules. Herein, we describe a survey calculation for electronic states with the leading configuration containing two Th $7s$ electrons and an unpaired electron populated in a high-lying orbital, hereafter referred to as $s^2l$ states, to obtain a qualitative understanding of their properties as well as a rough idea of their term energies. The electron-attachment version of the \emph{equation-of-motion coupled-cluster singles and doubles} method \cite{EOMEA} has been used with the $7s^2$ configuration of ThF$^+$ as the reference state. An additional electron is then attached to form the target $s^2l$ states of ThF molecule. The SFX2C-1e scheme \cite{SFX2C1e,Cheng2011} has been used for treating relativistic effects. The backbone of the basis set for Th was formed by taking the $s$-, $p$-, $d$-, and $f$-type functions in the ANO-RCC set in the fully uncontracted fashion together with $g$- and $h$-type functions from the cc-pwCVTZ-X2C basis set \cite{Th-pwcvtz}. In addition, three sets of additional diffuse $s$-, $p$-, $d$-, and $f$-type functions generated using a geometric factor of 3.0 have been included in order to capture the possible Rydberg nature in the excited state wavefunctions. The resulting basis set for Th has the pattern $29s26p22d18f4g1h$. The standard aug-cc-pVTZ basis set for F has been used in the uncontracted form. The computational results are summarized in Table \ref{tab:eomccsd}. 
        \begin{table}[htb]
            \centering
            \begin{tabular}{| c | c c c c |}
                \hline
                & $T_e$ & Osc. Str. & Unpaired electron wavefunction & $B_e$ \\
                \hline
                $X ^2\Delta $ & 0 & - & $6d(x^2-y^2)$, $6d(xy)$ & 0.233 \\
                $1 ^2\Pi $ & 3173 & - & $6d(xz)$, $6d(yz)$ & 0.225 \\
                $1 ^2\Sigma $ & 5905 & - & $6d(z^2)$ & 0.226 \\
                $1 ^2\Phi $ & 11183 & 0.05 & $5f(x^2y)$, $5f(xy^2)$ & 0.225 \\
                $2 ^2\Delta $ & 11318 & 0.24 & $5f(xyz)$, $5f(x^2z-y^2z)$ & 0.227 \\
                $2 ^2\Pi $ & 13818 & 1 & $7p(x)$, $7p(y)$ & 0.232 \\
                $2 ^2\Sigma $ & 17947 & - & $7p(z)$ & 0.223 \\
                $3 ^2\Pi $ & 22745 & 1.01 & $5f(z^2x)+8p(x)$, $5f(z^2y)+8p(y)$ & 0.224 \\
                $3 ^2\Sigma $ & 33286 & - & $8s$ & 0.240 \\
                $4 ^2\Sigma $ & 35963 & - & $8p(z)$ & 0.238 \\
                $4 ^2\Pi $ & 38351 & 0.06 & $8p(x)+7d(xz)$, $8p(y)+7d(yz)$ & 0.244 \\
                $3 ^2\Delta $ & 38476 & 0.39 & $7d(xy)$, $7d(x^2-y^2)$ & 0.242 \\
                $5 ^2\Sigma $ & 38604 & 0 & $7d(z^2)$ & 0.245 \\
                $5 ^2\Pi $ & 39009 & 0.06 & $7d(xz)$, $7d(yz)$ & 0.246 \\
                $6 ^2\Sigma $ & 41724 & - & $9s$ & 0.246 \\
                $4 ^2\Delta $ & 43418 & 0.51 & $6f(xyz)$, $6f(x^2z-y^2z)$ & 0.244 \\
                $2 ^2\Phi $ & 43459 & 3.11 & $6f(x^2y)$, $6f(xy^2)$ & 0.246 \\
                $5 ^2\Pi $ & 43618 & 1.33 & $6f(z^2x)$, $6f(z^2y)$ & 0.244 \\
                $5 ^2\Delta $ & 43731 & - & $8d(xy)$, $8d(x^2-y^2)$ & - \\
                $6 ^2\Pi $ & 43885 & - & $8d(xz)$, $8d(yz)$ & - \\
                \hline
                IP & 50586 & - & - & 0.247 \\
                \hline
            \end{tabular}
            \caption{Molecular constants of core-nonpenetrating Rydberg states of ThF$^+$ calculated with the \emph{equation-of-motion coupled-cluster singles and doubles} method. Values for $T_e$ (term energy) and $B_e$ (equilibrium rotational constant) are given in units of cm$^{-1}$. Rotational constants of these states are large at high energies, allowing us to distinguish between the $s^2f$ and $spd$ states from the rotational constant alone. Unfortunately, our spectra between 36000 and 44000~cm$^{-1}$ do not reveal any bands with rotational constants larger than 0.225~cm$^{-1}$.}
            \label{tab:eomccsd}
        \end{table}
        
        Referring to Table \ref{tab:eomccsd}, the $7s^27d$ and $7s^26f$ states lie at around 38000 to 43000~cm$^{-1}$, which is easily accessible by an ultraviolet pulsed laser. A prominent feature of these states is that the rotational constants ($B=$ 0.242 and 0.246~cm$^{-1}$, respectively) are substantially larger than that of the ground state (0.232~cm$^{-1}$). This can be qualitatively understood as enhanced attraction between the partially positively-charged Th center with the partially negatively-charged F center, because the Rydberg electron has been excited from its ground state into a diffuse (very weakly shielding) orbital. This interpretation is also consistent with the fact that the $B$ constants of these excited states are close to that of the ionized state (0.247~cm$^{-1}$). 

        Based on the theoretical calculations, one experimentally significant difference between the $s^2f$ and the $spd$ states lies in their rotational constants, especially for states with high excitation energy. Hence, we should be able to distinguish between the $s^2f$ and $spd$ states by looking at their rotational constants. Unfortunately, our spectra between 36000 and 44000~cm$^{-1}$ do not reveal any bands with rotational constants larger than 0.225~cm$^{-1}$. To make things worse, instead of observing regular and strong transitions within the target region suggested by theoretical predictions above 38000~cm$^{-1}$, we observe weak and irregular clusters of transitions. We cannot tell whether any of these transitions terminate in the $s^2l$ states. The weak and irregular nature of these excited states is tentatively explained in the following section. 

    \subsection{Perturbations in the deep ultraviolet region}\label{sec:perturbationDeepUV}
        Our explanation for the lack of regular rotational band structures in the deep ultraviolet region is the result of perturbations between the $s^2f$ states and $spd$ states. In contrast to the region where only $spd$ states exist, i.e.\ low lying states, the diabatic potential curves of both types of states will cross very frequently at high energy levels, because of significantly different internuclear distances (different rotational constants of $spd$ and $s^2f$). In this region, interactions could give rise to many avoided crossings, and profoundly distort the adiabatic potentials for both the $s^2f$ and $spd$ states. The crossings of these potential curves would redistribute the oscillator strength of a vibrational level in a diabatic potential to many other vibrational levels of other potentials, hence making the transitions very weak and with an irregular vibrational pattern. 
        %as illustrated in Figure \ref{fig:differentB}.
        %\begin{figure}[htb]
        %    \centering
        %    \begin{subfigure}[htb]{0.47\columnwidth}
        %        \includegraphics[width=\textwidth]{similarB.pdf}
        %        \caption{Similar internuclear distances}
        %        \label{fig:similarB}
        %    \end{subfigure}
        %    \vspace{5pt}
        %    \begin{subfigure}[htb]{0.47\columnwidth}
        %        \includegraphics[width=\textwidth]{differentB.pdf}
        %        \caption{Different internuclear distances}
        %        \label{fig:differentB}
        %    \end{subfigure}            
        %    \caption{Crossings of diabatic potential curves of $spd$ and $s^2l$ states distort the potential curves. The horizontal lines represent the vibrational levels. Left: low lying $spd$ and $s^2f$ states have similar internuclear distances, so their potential curves do not cross each other. Right: high lying $spd$ and $s^2f$ states have different internuclear distances, so their potential curves (dashed) cross each other and mix, giving rise to very different adiabatic potential curves (solid). }
        %    \label{fig:perturbationCrossing}
        %\end{figure}
        
        Testing this hypothesis quantitatively would require a significant amount of high resolution spectra in this region. In addition, theoretical calculations of 2-electron or multiple 1-electron spin-orbit perturbations, such as $spd \sim s^2d \sim s^2f$ states are also required. These fundamental studies are beyond the scope of this paper.

\section{Conclusion}
    We have made extensive observations of the spectra of ThF and ThO, using LIF and REMPI spectroscopy on a supersonic cooled molecular beam formed by laser ablation. We have recorded 345 ThF vibronic bands between 13800 and 44600~cm$^{-1}$. Among these bands, we have identified 8 sets of ThF vibrational progressions by performing a combination difference analysis. Since all of the identified transitions are from the electronic ground state, we perform a global fit to improve the precision of the molecular constants of the ground state of ThF from previous works. A high precision \emph{ab initio} calculation has also been performed for comparison with experimental measurements. 

    The two-photon resonance enhanced photoionization scheme that we use results in 30\% of ThF$^+$ ions being produced in the the eEDM sensitive state ($^3\Delta_1$, $v^+=0$, $J^+=1$). These ions will be loaded into a RF Paul trap for the eEDM precision measurements. In addition, we propose a scheme of rotationally selective photoionization to a single rovibronic state of ThF$^+$. However, we have not been successful in identifying an appropriate intermediate state for our proposed excitation scheme, because of spectroscopic complexity in the deep ultraviolet region. Further theoretical calculations of interactions between $s^2f$ and $spd$ states are in progress.

\clearpage
\section*{Acknowledgement}
    Funding was provided by the Marsico Foundation, NIST, and the NSF Physics Frontier Center at JILA (PHY-1734006). K. B. Ng acknowledges support from the Tan Kah Kee Foundation in Singapore. L. Cheng thanks Johns Hopkins University for the start-up fund. R. W. Field thanks NSF grant (CHE-1800410). We thank Y. Shagam, W. Cairncross, T. Roussy, K. Boyce, and A. Vigil for useful discussions. Commercial products referenced in this work are not endorsed by NIST and are for the purposes of technical communication only.

\section*{References}

\bibliography{mybibfile}

\clearpage
\appendix
\setcounter{table}{0}
\setcounter{figure}{0}

\section{All ThF fitted transitions}
    The molecular constants of all ThF transitions that we have fitted are shown in Tables \ref{tab:all1} to \ref{tab:all14}. We fix $B'' = 0.2325~\text{cm}^{-1}$, i.e.,\ the average of the rotational constant of the $v''=0$ and $v''=1$ of the ThF $^2\Delta_{3/2}$ ground electronic state, unless we assign that transition as a member of a vibrational progression. In that case, we use the $B''$ and $B'$ of the relevant lower and upper states of the transitions (see Section \ref{sec:family}). The values of the first pair of parentheses of $B'$ are fitting errors, and those in the second pair of parentheses are uncertainties from unknown initial vibrational states. The fitting error bars are quoted to 90\% confidence in the fit. We abbreviate $(p+2q)/2$ as $p2q$.
    \begin{table}[htb]
        \centering
        \begin{tabular}{| c c c c | c | c |}
        \hline
        $\nu_0$ & $B'$ & $B''$ & $\Omega'$ & $v'-v''$ & Additional comments \\
        \hline
 \text{13819.21(2)} & \text{0.2173(3)(7)} & 0.2325 & \text{3/2} & - & $p2q=0.0279(16)$ \\
 \text{13871.33(1)} & \text{0.2248(2)(7)} & 0.2325 & \text{3/2} & - & - \\
 \text{13927.97(1)} & \text{0.2196(2)(7)} & 0.2325 & \text{5/2} & - & - \\
 \text{13939.69(1)} & \text{0.2244(2)(7)} & 0.2325 & \text{3/2} & - & - \\
 \text{13978.07(2)} & \text{0.2210(3)} & 0.2290 & \text{3/2} & $1'-3''$ & $\Omega=3/2$ [15.18] \\
 \text{13982.65(2)} & \text{0.2221(3)} & 0.2304 & \text{3/2} & $0'-2''$ & $\Omega=3/2$ [15.18] \\
 \text{14008.752(9)} & \text{0.2242(1)(7)} & 0.2325 & \text{3/2} & - & - \\
 \text{14032.84(3)} & \text{0.2248(4)} & 0.2318 & \text{3/2} & $1'-1''$ & $\Omega=3/2$ [14.09] \\
 \text{14072.447(8)} & \text{0.2267(2)} & 0.2332 & \text{3/2} & $0'-0''$ & $\Omega=3/2$ [14.09] \\
 \text{14198.18(2)} & \text{0.2237(3)} & 0.2304 & \text{3/2} & $?'-2''$ & - \\
 \text{14202.35(1)} & \text{0.2229(2)(7)} & 0.2325 & \text{3/2} & - & - \\
 \text{14282.18(2)} & \text{0.2247(2)(7)} & 0.2325 & \text{3/2} & - & - \\
 \text{14566.67(1)} & \text{0.2217(2)} & 0.2304 & \text{3/2} & $1'-2''$ & $\Omega=3/2$ [15.18] \\
 \text{14575.37(1)} & \text{0.2223(2)} & 0.2318 & \text{3/2} & $0'-1''$ & $\Omega=3/2$ [15.18] \\
 \text{14629.699(9)} & \text{0.2250(1)} & 0.2332 & \text{3/2} & $1'-0''$ & $\Omega=3/2$ [14.09] \\
 \text{14685.52(2)} & \text{0.2217(3)(7)} & 0.2325 & \text{3/2} & - & - \\
        \hline
        \end{tabular}
        \caption{Molecular parameters of all ThF transitions fitted. Units for $\nu_0$, $B'$, and $B''$ are in cm$^{-1}$.}
        \label{tab:all1}
    \end{table}
    
    \begin{table}[htb]
        \centering
        \begin{tabular}{| c c c c | c | c |}
        \hline
        $\nu_0$ & $B'$ & $B''$ & $\Omega'$ & $v'-v''$ & Additional comments \\
        \hline
 \text{14745.70(3)} & \text{0.2221(3)(7)} & 0.2325 & \text{3/2} & - & - \\
 \text{14790.89(1)} & \text{0.2242(2)} & 0.2318 & \text{3/2} & $?'-1''$ & - \\
 \text{14836.92(1)} & \text{0.2225(1)(7)} & 0.2325 & \text{3/2} & - & - \\
 \text{15016.93(2)} & \text{0.2201(3)(7)} & 0.2325 & \text{3/2} & - & - \\
 \text{15061.42(2)} & \text{0.2171(4)} & 0.2290 & \text{3/2} & $?'-3''$ & - \\
 \text{15096.57(2)} & \text{0.2201(2)} & 0.2318 & \text{3/2} & $?'-1''$ & - \\
 \text{15123.43(2)} & \text{0.2185(3)(7)} & 0.2325 & \text{3/2} & - & - \\
 \text{15159.41(3)} & \text{0.2221(4)} & 0.2318 & \text{3/2} & $1'-1''$ & $\Omega=3/2$ [15.18] \\
 \text{15172.20(6)} & \text{0.223(1)} & 0.2332 & \text{3/2} & $0'-0''$ & $\Omega=3/2$ [15.18] \\
 \text{15225.60(3)} & \text{0.2223(4)(7)} & 0.2325 & \text{3/2} & - & - \\
 \text{15387.75(2)} & \text{0.2246(4)} & 0.2332 & \text{3/2} & $?'-0''$ & - \\
 \text{15450.56(3)} & \text{0.2193(5)(7)} & 0.2325 & \text{3/2} & - & - \\
 \text{15471.49(3)} & \text{0.2186(5)(7)} & 0.2325 & \text{-} & - & $\gamma=-0.2589(50)$, $\Lambda=1$ \\
 \text{15497.46(2)} & \text{0.2208(4)(7)} & 0.2325 & \text{-} & - & $\gamma=-0.1971(34)$, $\Lambda=1$ \\
 \text{15650.11(3)} & \text{0.2164(5)} & 0.2304 & \text{3/2} & $?'-2''$ & - \\
 \text{15693.44(5)} & \text{0.2203(7)} & 0.2332 & \text{3/2} & $?'-0''$ & - \\
 \text{15756.25(2)} & \text{0.2222(4)} & 0.2332 & \text{3/2} & $1'-0''$ & $\Omega=3/2$ [15.18] \\
 \text{15766.61(3)} & \text{0.2238(5)(7)} & 0.2325 & \text{3/2} & - & - \\
 \text{15865.75(2)} & \text{0.2241(4)(7)} & 0.2325 & \text{3/2} & - & - \\
 \text{15932.71(2)} & \text{0.2236(3)(7)} & 0.2325 & \text{3/2} & - & - \\
 \text{16034.59(2)} & \text{0.2204(3)(7)} & 0.2325 & \text{-} & - & $\Lambda=0$ \\
 \text{16126.25(2)} & \text{0.2211(3)(7)} & 0.2325 & \text{-} & - & $\gamma=-0.3180(34)$, $\Lambda=0$ \\
 \text{16275.97(2)} & \text{0.2221(4)(7)} & 0.2325 & \text{5/2} & - & - \\
 \text{16363.48(2)} & \text{0.2224(4)(7)} & 0.2325 & \text{3/2} & - & - \\
 \text{18056.78(2)} & \text{0.2180(3)(7)} & 0.2325 & \text{3/2} & - & - \\
 \text{18323.77(4)} & \text{0.2207(7)} & 0.2318 & \text{3/2} & $?'-1''$ & - \\
 \text{18491.18(2)} & \text{0.2160(4)} & 0.2318 & \text{3/2} & $1'-1''$ & $\Omega=3/2$ [18.62] \\
 \text{18574.48(2)} & \text{0.2172(3)} & 0.2332 & \text{3/2} & $0'-0''$ & $\Omega=3/2$ [18.62] \\
 \text{18920.61(2)} & \text{0.2210(4)} & 0.2332 & \text{3/2} & $?'-0''$ & - \\
        \hline
        \end{tabular}
        \caption{Molecular parameters of all ThF transitions fitted. Units for $\nu_0$, $B'$, and $B''$ are in cm$^{-1}$.}
        \label{tab:all2}
    \end{table}
    
    \begin{table}[htb]
        \centering
        \begin{tabular}{| c c c c | c | c |}
        \hline
        $\nu_0$ & $B'$ & $B''$ & $\Omega'$ & $v'-v''$ & Additional comments \\
        \hline
 \text{19000.73(2)} & \text{0.2147(2)} & 0.2318 & \text{3/2} & $2'-1''$ & $\Omega=3/2$ [18.62] \\
 \text{19088.03(2)} & \text{0.2160(2)} & 0.2332 & \text{3/2} & $1'-0''$ & $\Omega=3/2$ [18.62] \\
 \text{19185.03(2)} & \text{0.2196(4)(7)} & 0.2325 & \text{1/2} & - & $p2q=0.0586(17)$ \\
 \text{19288.18(3)} & \text{0.2162(4)} & 0.2304 & \text{3/2} & $1'-2''$ & $\Omega=3/2$ [19.98] \\
 \text{19347.86(2)} & \text{0.2169(2)} & 0.2318 & \text{3/2} & $0'-1''$ & $\Omega=3/2$ [19.98] \\
 \text{19399.31(2)} & \text{0.2155(5)(7)} & 0.2325 & \text{5/2} & - & - \\
 \text{19418.70(4)} & \text{0.2271(8)(7)} & 0.2325 & \text{1/2} & - & - \\
 \text{19457.01(2)} & \text{0.2194(3)(7)} & 0.2325 & \text{3/2} & - & - \\
 \text{19506.29(2)} & \text{0.2144(2)(7)} & 0.2325 & \text{3/2} & - & - \\
 \text{19737.19(4)} & \text{0.2179(8)(7)} & 0.2325 & \text{3/2} & - & - \\
 \text{19762.05(2)} & \text{0.2150(3)} & 0.2304 & \text{3/2} & $1'-2''$ & $\Omega=3/2$ [20.40] \\
 \text{19785.30(2)} & \text{0.2135(2)} & 0.2318 & \text{3/2} & $0'-1''$ & $\Omega=3/2$ [20.40] \\
 \text{19829.45(1)} & \text{0.2176(2)(7)} & 0.2325 & \text{3/2} & - & - \\
 \text{19846.98(1)} & \text{0.2167(2)(7)} & 0.2325 & \text{3/2} & - & - \\
 \text{19880.91(1)} & \text{0.2163(2)} & 0.2318 & \text{3/2} & $1'-1''$ & $\Omega=3/2$ [19.98] \\
 \text{19924.61(1)} & \text{0.2202(2)(7)} & 0.2325 & \text{3/2} & - & - \\
 \text{19944.72(1)} & \text{0.2169(2)} & 0.2332 & \text{3/2} & $0'-0''$ & $\Omega=3/2$ [19.98] \\
 \text{19977.73(1)} & \text{0.2175(2)(7)} & 0.2325 & \text{3/2} & - & - \\
 \text{20237.35(2)} & \text{0.2179(3)(7)} & 0.2325 & \text{3/2} & - & - \\
 \text{20289.16(2)} & \text{0.2197(4)(7)} & 0.2325 & \text{1/2} & - & $p2q=0.0189(14)$ \\
 \text{20354.77(1)} & \text{0.2154(2)} & 0.2318 & \text{3/2} & $1'-1''$ & $\Omega=3/2$ [20.40] \\
 \text{27553.79(2)} & \text{0.2120(4)(7)} & 0.2325 & \text{5/2} & - & - \\
 \text{27575.88(1)} & \text{0.2141(2)(7)} & 0.2325 & \text{3/2} & - & - \\
 \text{27592.24(2)} & \text{0.2113(3)(7)} & 0.2325 & \text{5/2} & - & - \\
 \text{27762.49(2)} & \text{0.2140(3)} & 0.2318 & \text{3/2} & $0'-1''$ & $\Omega=3/2$ [28.37] \\
 \text{27814.94(2)} & \text{0.2074(3)(7)} & 0.2325 & \text{3/2} & - & - \\
 \text{27858.22(3)} & \text{0.2147(4)(7)} & 0.2325 & \text{3/2} & - & - \\
 \text{27890.49(2)} & \text{0.2117(5)(7)} & 0.2325 & \text{3/2} & - & - \\
        \hline
        \end{tabular}
        \caption{Molecular parameters of all ThF transitions fitted. Units for $\nu_0$, $B'$, and $B''$ are in cm$^{-1}$.}
        \label{tab:all3}
    \end{table}
    
    \begin{table}[htb]
        \centering
        \begin{tabular}{| c c c c | c | c |}
        \hline
        $\nu_0$ & $B'$ & $B''$ & $\Omega'$ & $v'-v''$ & Additional comments \\
        \hline
 \text{27925.77(2)} & \text{0.2159(4)} & 0.2318 & \text{1/2} & $1'-1''$ & $p2q=0.0074(16)$, $^2\Pi_{1/2}$ [27.98]\\
 \text{27959.51(2)} & \text{0.2169(4)} & 0.2332 & \text{1/2} & $0'-0''$ & $p2q=0.0094(15)$, $^2\Pi_{1/2}$ [27.98]\\
 \text{27976.03(1)} & \text{0.2203(2)(7)} & 0.2325 & \text{3/2} & - & - \\
 \text{27990.31(3)} & \text{0.2133(5)(7)} & 0.2325 & \text{5/2} & - & - \\
 \text{27997.16(1)} & \text{0.2178(2)(7)} & 0.2325 & \text{1/2} & - & $p2q=0.0632(12)$ \\
 \text{28014.97(2)} & \text{0.2164(3)(7)} & 0.2325 & \text{5/2} & - & - \\
 \text{28088.25(3)} & \text{0.2143(4)(7)} & 0.2325 & \text{1/2} & - & $p2q=0.0460(19)$ \\
 \text{28194.75(1)} & \text{0.2141(2)(7)} & 0.2325 & \text{1/2} & - & $p2q=0.0251(10)$ \\
 \text{28231.14(2)} & \text{0.2151(3)(7)} & 0.2325 & \text{5/2} & - & - \\
 \text{28262.09(1)} & \text{0.2169(2)(7)} & 0.2325 & \text{1/2} & - & $p2q=0.0236(10)$ \\
 \text{28271.20(2)} & \text{0.2130(3)(7)} & 0.2325 & \text{3/2} & - & - \\
 \text{28321.52(3)} & \text{0.2085(4)(7)} & 0.2325 & \text{5/2} & - & - \\
 \text{28342.89(2)} & \text{0.2151(3)(7)} & 0.2325 & \text{1/2} & - & $p2q=0.0241(12)$ \\
 \text{28359.34(3)} & \text{0.2142(4)} & 0.2332 & \text{3/2} & $0'-0''$ & $\Omega=3/2$ [28.37] \\
 \text{28377.77(2)} & \text{0.2128(4)(7)} & 0.2325 & \text{5/2} & - & - \\
 \text{28416.67(2)} & \text{0.2151(3)(7)} & 0.2325 & \text{1/2} & - & $p2q=0.0258(17)$ \\
 \text{28435.00(2)} & \text{0.2120(3)(7)} & 0.2325 & \text{3/2} & - & - \\
 \text{28442.47(2)} & \text{0.2181(4)(7)} & 0.2325 & \text{1/2} & - & - \\
 \text{28475.89(3)} & \text{0.2130(4)(7)} & 0.2325 & \text{5/2} & - & - \\
 \text{28492.03(2)} & \text{0.2127(4)(7)} & 0.2325 & \text{5/2} & - & - \\
 \text{28522.65(3)} & \text{0.2164(7)} & 0.2318 & \text{1/2} & $1'-0''$ & $p2q=0.0110(30)$, $^2\Pi_{1/2}$ [27.98] \\
 \text{28615.73(2)} & \text{0.2176(3)} & 0.2318 & \text{1/2} & $0'-1''$ & $p2q=0.0203(14)$, $^2\Pi_{1/2}$ [29.22] \\
 \text{28670.79(1)} & \text{0.2194(2)(7)} & 0.2325 & \text{1/2} & - & $p2q=0.0296(12)$ \\
 \text{28699.29(2)} & \text{0.2148(4)(7)} & 0.2325 & \text{5/2} & - & - \\
 \text{28743.01(1)} & \text{0.2153(2)(7)} & 0.2325 & \text{1/2} & - & - \\
 \text{28747.54(4)} & \text{0.2102(7)(7)} & 0.2325 & \text{5/2} & - & - \\
 \text{28803.71(4)} & \text{0.2157(6)(7)} & 0.2325 & \text{3/2} & - & - \\
        \hline
        \end{tabular}
        \caption{Molecular parameters of all ThF transitions fitted. Units for $\nu_0$, $B'$, and $B''$ are in cm$^{-1}$.}
        \label{tab:all4}
    \end{table}
    
    \begin{table}[htb]
        \centering
        \begin{tabular}{| c c c c | c | c |}
        \hline
        $\nu_0$ & $B'$ & $B''$ & $\Omega'$ & $v'-v''$ & Additional comments \\
        \hline
 \text{28826.80(5)} & \text{0.2159(8)} & 0.2318 & \text{5/2} & $?'-1''$ & - \\
 \text{28853.26(2)} & \text{0.2153(3)(7)} & 0.2325 & \text{1/2} & - & $p2q=0.0280(13)$ \\
 \text{28868.24(2)} & \text{0.2107(3)(7)} & 0.2325 & \text{5/2} & - & - \\
 \text{28936.94(1)} & \text{0.2147(2)} & 0.2332 & \text{3/2} & $1'-0''$ & $\Omega=3/2$ [28.37] \\
 \text{28974.65(2)} & \text{0.2108(3)(7)} & 0.2325 & \text{5/2} & - & - \\
 \text{29056.14(2)} & \text{0.2143(3)(7)} & 0.2325 & \text{5/2} & - & - \\
 \text{29151.31(2)} & \text{0.2128(3)} & 0.2318 & \text{5/2} & - & - \\
 \text{29165.71(2)} & \text{0.2106(3)(7)} & 0.2325 & \text{5/2} & - & - \\
 \text{29169.54(2)} & \text{0.2072(5)(7)} & 0.2325 & \text{5/2} & - & - \\
 \text{29212.55(1)} & \text{0.2175(2)} & 0.2318 & \text{1/2} & $0'-0''$ & $p2q=0.01987(96)$, $^2\Pi_{1/2}$ [29.22] \\
 \text{29233.92(2)} & \text{0.2156(4)(7)} & 0.2325 & \text{3/2} & - & - \\
 \text{29255.92(3)} & \text{0.2098(5)(7)} & 0.2325 & \text{5/2} & - & - \\
 \text{29278.00(3)} & \text{0.2108(4)(7)} & 0.2325 & \text{3/2} & - & - \\
 \text{29334.73(2)} & \text{0.2219(5)(7)} & 0.2325 & \text{1/2} & - & $p2q=0.0162(19)$ \\
 \text{29349.34(2)} & \text{0.2098(2)(7)} & 0.2325 & \text{5/2} & - & - \\
 \text{29369.22(2)} & \text{0.2138(3)(7)} & 0.2325 & \text{5/2} & - & - \\
 \text{29423.62(3)} & \text{0.2168(5)} & 0.2332 & \text{5/2} & $?'-0''$ & - \\
 \text{29449.04(2)} & \text{0.2106(3)} & 0.2332 & \text{5/2} & - & - \\
 \text{29491.27(3)} & \text{0.2101(4)(7)} & 0.2325 & \text{1/2} & - & $p2q=0.0118(17)$ \\
 \text{29497.93(2)} & \text{0.2095(4)(7)} & 0.2325 & \text{5/2} & - & - \\
 \text{29502.04(3)} & \text{0.2201(5)(7)} & 0.2325 & \text{5/2} & - & - \\
 \text{29510.65(2)} & \text{0.2141(3)(7)} & 0.2325 & \text{5/2} & - & - \\
 \text{29542.11(2)} & \text{0.2165(3)(7)} & 0.2325 & \text{5/2} & - & - \\
 \text{29550.66(3)} & \text{0.2148(4)(7)} & 0.2325 & \text{5/2} & - & - \\
 \text{29582.46(2)} & \text{0.2155(4)(7)} & 0.2325 & \text{5/2} & - & - \\
        \hline
        \end{tabular}
        \caption{Molecular parameters of all ThF transitions fitted. Units for $\nu_0$, $B'$, and $B''$ are in cm$^{-1}$.}
        \label{tab:all5}
    \end{table}
    
    \begin{table}[htb]
        \centering
        \begin{tabular}{| c c c c | c | c |}
        \hline
        $\nu_0$ & $B'$ & $B''$ & $\Omega'$ & $v'-v''$ & Additional comments \\
        \hline
 \text{29596.08(2)} & \text{0.2154(3)(7)} & 0.2325 & \text{1/2} & - & $p2q=0.0241(15)$ \\
 \text{29601.51(4)} & \text{0.2110(5)(7)} & 0.2325 & \text{3/2} & - & - \\
 \text{29617.49(1)} & \text{0.2146(2)(7)} & 0.2325 & \text{5/2} & - & - \\
 \text{29626.63(4)} & \text{0.2154(5)(7)} & 0.2325 & \text{1/2} & - & $p2q=0.0263(26)$ \\
 \text{29663.17(2)} & \text{0.2236(3)(7)} & 0.2325 & \text{5/2} & - & - \\
 \text{29677.69(1)} & \text{0.2218(2)(7)} & 0.2325 & \text{1/2} & - & - \\
 \text{29681.34(4)} & \text{0.2098(6)} & 0.2318 & \text{5/2} & $?'-2''$ & - \\
 \text{29696.73(2)} & \text{0.2166(3)(7)} & 0.2325 & \text{1/2} & - & $p2q=0.0566(16)$ \\
 \text{29725.76(2)} & \text{0.2200(3)(7)} & 0.2325 & \text{5/2} & - & - \\
 \text{29748.33(3)} & \text{0.2128(5)(7)} & 0.2325 & \text{5/2} & - & - \\
 \text{29753.91(2)} & \text{0.2173(5)} & 0.2318 & \text{1/2} & $2'-1''$ & $p2q=0.0214(21)$, $^2\Pi_{1/2}$ [29.22] \\
 \text{29772.40(1)} & \text{0.2153(3)(7)} & 0.2325 & \text{5/2} & - & - \\
 \text{29785.48(2)} & \text{0.2177(3)} & 0.2332 & \text{1/2} & $1'-0''$ & $p2q=0.0237(14)$, $^2\Pi_{1/2}$ [29.22] \\
 \text{29801.72(2)} & \text{0.2080(3)} & 0.2318 & \text{5/2} & $?'-1''$ & - \\
 \text{29812.09(2)} & \text{0.2158(4)(7)} & 0.2325 & \text{1/2} & - & $p2q=0.0077(14)$ \\
 \text{29824.86(1)} & \text{0.2153(2)(7)} & 0.2325 & \text{1/2} & - & $p2q=0.0284(10)$ \\
 \text{29837.72(3)} & \text{0.2068(5)(7)} & 0.2325 & \text{5/2} & - & - \\
 \text{29842.62(3)} & \text{0.2094(5)} & 0.2318 & \text{5/2} & $?'-1''$ & - \\
 \text{29889.90(1)} & \text{0.2099(5)} & 0.2318 & \text{1/2} & $?'-1''$ & - \\
 \text{29903.07(4)} & \text{0.2072(6)(7)} & 0.2325 & \text{5/2} & - & - \\
 \text{29910.38(3)} & \text{0.2107(4)(7)} & 0.2325 & \text{5/2} & - & - \\
 \text{29917.54(2)} & \text{0.2025(3)(7)} & 0.2325 & \text{5/2} & - & - \\
 \text{29928.02(1)} & \text{0.2135(2)(7)} & 0.2325 & \text{5/2} & - & - \\
 \text{29932.46(3)} & \text{0.2008(6)(7)} & 0.2325 & \text{5/2} & - & - \\
 \text{29935.72(2)} & \text{0.2158(4)} & 0.2318 & \text{5/2} & $?'-1''$ & - \\
        \hline
        \end{tabular}
        \caption{Molecular parameters of all ThF transitions fitted. Units for $\nu_0$, $B'$, and $B''$ are in cm$^{-1}$.}
        \label{tab:all6}
    \end{table}
    
    \begin{table}[htb]
        \centering
        \begin{tabular}{| c c c c | c | c |}
        \hline
        $\nu_0$ & $B'$ & $B''$ & $\Omega'$ & $v'-v''$ & Additional comments \\
        \hline
 \text{29959.28(2)} & \text{0.2099(3)(7)} & 0.2325 & \text{3/2} & - & - \\
 \text{29974.26(5)} & \text{0.2099(9)(7)} & 0.2325 & \text{5/2} & - & - \\
 \text{30021.07(2)} & \text{0.2167(3)(7)} & 0.2325 & \text{1/2} & - & - \\
 \text{30030.21(4)} & \text{0.2108(5)(7)} & 0.2325 & \text{3/2} & - & - \\
 \text{30046.06(9)} & \text{0.2144(13)(7)} & 0.2325 & \text{1/2} & - & $p2q=0.0453(57)$ \\
 \text{30049.38(3)} & \text{0.2148(7)(7)} & 0.2325 & \text{5/2} & - & - \\
 \text{30074.85(3)} & \text{0.2158(4)(7)} & 0.2325 & \text{1/2} & - & - \\
 \text{30093.56(3)} & \text{0.2192(11)(7)} & 0.2325 & \text{1/2} & - & $p2q=0.0394(60)$ \\
 \text{30111.14(2)} & \text{0.2158(6)(7)} & 0.2325 & \text{1/2} & - & $p2q=0.0609(20)$ \\
 \text{30124.07(2)} & \text{0.2149(5)(7)} & 0.2325 & \text{1/2} & - & - \\
 \text{30147.28(2)} & \text{0.2158(7)(7)} & 0.2325 & \text{1/2} & - & - \\
 \text{30159.17(1)} & \text{0.2135(5)} & 0.2318 & \text{5/2} & $?'-1''$ & - \\
 \text{30169.68(6)} & \text{0.2133(9)(7)} & 0.2325 & \text{5/2} & - & - \\
 \text{30229.11(2)} & \text{0.2104(7)(7)} & 0.2325 & \text{5/2} & - & - \\
 \text{30239.27(3)} & \text{0.2085(9)(7)} & 0.2325 & \text{5/2} & - & - \\
 \text{30240.81(4)} & \text{0.2073(10)(7)} & 0.2325 & \text{5/2} & - & - \\
 \text{30274.04(2)} & \text{0.2096(3)} & 0.2318 & \text{5/2} & $?'-1''$ & - \\
 \text{30300.96(2)} & \text{0.2153(6)(7)} & 0.2325 & \text{1/2} & - & - \\
 \text{30342.73(1)} & \text{0.2088(4)(7)} & 0.2325 & \text{5/2} & - & - \\
 \text{30350.89(3)} & \text{0.2174(4)} & 0.2332 & \text{1/2} & $2'-0''$ & $p2q=0.0232(21)$, $^2\Pi_{1/2}$ [29.22] \\
 \text{30353.94(2)} & \text{0.2119(8)(7)} & 0.2325 & \text{5/2} & - & - \\
 \text{30370.36(2)} & \text{0.2053(6)(7)} & 0.2325 & \text{1/2} & - & - \\
 \text{30381.66(2)} & \text{0.2130(4)(7)} & 0.2325 & \text{1/2} & - & $p2q=0.0165(23)$ \\
 \text{30387.74(3)} & \text{0.2115(8)(7)} & 0.2325 & \text{3/2} & - & - \\
 \text{30398.58(3)} & \text{0.2080(4)} & 0.2332 & \text{5/2} & $?'-0''$ & - \\
        \hline
        \end{tabular}
        \caption{Molecular parameters of all ThF transitions fitted. Units for $\nu_0$, $B'$, and $B''$ are in cm$^{-1}$.}
        \label{tab:all7}
    \end{table}
    
    \begin{table}[htb]
        \centering
        \begin{tabular}{| c c c c | c | c |}
        \hline
        $\nu_0$ & $B'$ & $B''$ & $\Omega'$ & $v'-v''$ & Additional comments \\
        \hline
 \text{30409.62(4)} & \text{0.2112(13)(7)} & 0.2325 & \text{5/2} & - & - \\
 \text{30416.11(2)} & \text{0.2169(7)} & 0.2318 & \text{3/2} & $?'-1''$ & - \\
 \text{30430.00(3)} & \text{0.2115(8)(7)} & 0.2325 & \text{5/2} & - & - \\
 \text{30439.45(2)} & \text{0.2103(2)} & 0.2332 & \text{5/2} & $?'-0''$ & - \\
 \text{30445.47(2)} & \text{0.2086(4)(7)} & 0.2325 & \text{5/2} & - & - \\
 \text{30450.48(2)} & \text{0.2150(5)(7)} & 0.2325 & \text{3/2} & - & - \\
 \text{30486.79(2)} & \text{0.2096(8)} & 0.2332 & \text{1/2} & $?'-0''$ & - \\
 \text{30494.42(3)} & \text{0.2138(13)(7)} & 0.2325 & \text{3/2} & - & - \\
 \text{30500.84(4)} & \text{0.2139(11)(7)} & 0.2325 & \text{1/2} & - & $p2q=0.0458(43)$ \\
 \text{30523.33(3)} & \text{0.2094(7)(7)} & 0.2325 & \text{3/2} & - & - \\
 \text{30532.51(2)} & \text{0.2199(8)} & 0.2332 & \text{5/2} & $?'-0''$ & - \\
 \text{30540.52(2)} & \text{0.2116(8)(7)} & 0.2325 & \text{1/2} & - & - \\
 \text{30585.27(6)} & \text{0.2159(18)(7)} & 0.2325 & \text{1/2} & - & - \\
 \text{30590.15(1)} & \text{0.2172(5)(7)} & 0.2325 & \text{1/2} & - & $p2q=0.0568(20)$ \\
 \text{30600.52(2)} & \text{0.2077(5)(7)} & 0.2325 & \text{3/2} & - & - \\
 \text{30616.20(3)} & \text{0.2022(8)(7)} & 0.2325 & \text{3/2} & - & - \\
 \text{30621.08(4)} & \text{0.2128(14)(7)} & 0.2325 & \text{5/2} & - & - \\
 \text{30628.58(2)} & \text{0.2146(6)(7)} & 0.2325 & \text{1/2} & - & - \\
 \text{30657.16(2)} & \text{0.2130(6)(7)} & 0.2325 & \text{5/2} & - & - \\
 \text{30665.10(2)} & \text{0.2135(6)(7)} & 0.2325 & \text{5/2} & - & - \\
 \text{30706.26(1)} & \text{0.2126(4)(7)} & 0.2325 & \text{3/2} & - & - \\
 \text{30756.04(2)} & \text{0.2137(4)} & 0.2332 & \text{5/2} & $?'-0''$ & - \\
 \text{30797.47(4)} & \text{0.2159(10)(7)} & 0.2325 & \text{1/2} & - & - \\
 \text{30826.38(2)} & \text{0.2160(4)(7)} & 0.2325 & \text{1/2} & - & - \\
 \text{30839.20(2)} & \text{0.2120(8)(7)} & 0.2325 & \text{5/2} & - & - \\
        \hline
        \end{tabular}
        \caption{Molecular parameters of all ThF transitions fitted. Units for $\nu_0$, $B'$, and $B''$ are in cm$^{-1}$.}
        \label{tab:all8}
    \end{table}
    
    \begin{table}[htb]
        \centering
        \begin{tabular}{| c c c c | c | c |}
        \hline
        $\nu_0$ & $B'$ & $B''$ & $\Omega'$ & $v'-v''$ & Additional comments \\
        \hline
 \text{30918.86(3)} & \text{0.2127(9)(7)} & 0.2325 & \text{5/2} & - & - \\
 \text{30933.79(3)} & \text{0.2099(9)(7)} & 0.2325 & \text{5/2} & - & - \\
 \text{30980.95(2)} & \text{0.2101(7)(7)} & 0.2325 & \text{3/2} & - & - \\
 \text{31012.93(2)} & \text{0.2176(6)} & 0.2332 & \text{3/2} & $?'-0''$ & - \\
 \text{31029.72(2)} & \text{0.2128(7)(7)} & 0.2325 & \text{1/2} & - & - \\
 \text{31051.68(2)} & \text{0.2120(7)(7)} & 0.2325 & \text{1/2} & - & - \\
 \text{31090.09(4)} & \text{0.2190(11)(7)} & 0.2325 & \text{3/2} & - & - \\
 \text{31095.00(3)} & \text{0.2088(11)(7)} & 0.2325 & \text{5/2} & - & - \\
 \text{31135.08(3)} & \text{0.2203(10)(7)} & 0.2325 & \text{5/2} & - & - \\
 \text{31146.81(3)} & \text{0.2151(11)} & 0.2318 & \text{5/2} & $?'-1''$ & - \\
 \text{31230.47(3)} & \text{0.2164(10)(7)} & 0.2325 & \text{3/2} & - & - \\
 \text{31381.88(2)} & \text{0.2085(5)(7)} & 0.2325 & \text{1/2} & - & - \\
 \text{31387.33(2)} & \text{0.2114(7)(7)} & 0.2325 & \text{1/2} & - & - \\
 \text{31478.17(3)} & \text{0.2110(9)(7)} & 0.2325 & \text{1/2} & - & - \\
 \text{31517.51(3)} & \text{0.2143(10)(7)} & 0.2325 & \text{3/2} & - & - \\
 \text{31530.57(3)} & \text{0.2156(8)(7)} & 0.2325 & \text{5/2} & - & - \\
 \text{31582.86(2)} & \text{0.2172(6)(7)} & 0.2325 & \text{5/2} & - & - \\
 \text{31685.36(2)} & \text{0.2159(7)(7)} & 0.2325 & \text{1/2} & - & $p2q=0.0408(26)$ \\
 \text{31698.92(2)} & \text{0.2148(6)(7)} & 0.2325 & \text{1/2} & - & $p2q=0.0393(27)$ \\
 \text{31723.43(3)} & \text{0.2118(9)(7)} & 0.2325 & \text{5/2} & - & - \\
 \text{31743.70(2)} & \text{0.2147(7)} & 0.2332 & \text{5/2} & $?'-0''$ & - \\
 \text{31784.43(3)} & \text{0.2097(8)(7)} & 0.2325 & \text{5/2} & - & - \\
 \text{31810.90(4)} & \text{0.2128(12)(7)} & 0.2325 & \text{5/2} & - & - \\
 \text{31964.14(2)} & \text{0.2136(4)} & 0.2318 & \text{5/2} & $?'-1''$ & - \\
 \text{31993.85(3)} & \text{0.2098(11)(7)} & 0.2325 & \text{5/2} & - & - \\
        \hline
        \end{tabular}
        \caption{Molecular parameters of all ThF transitions fitted. Units for $\nu_0$, $B'$, and $B''$ are in cm$^{-1}$.}
        \label{tab:all9}
    \end{table}

    \begin{table}[htb]
        \centering
        \begin{tabular}{| c c c c | c | c |}
        \hline
        $\nu_0$ & $B'$ & $B''$ & $\Omega'$ & $v'-v''$ & Additional comments \\
        \hline
 \text{32006.60(4)} & \text{0.2111(15)(7)} & 0.2325 & \text{5/2} & - & - \\
 \text{32023.46(1)} & \text{0.2156(3)(7)} & 0.2325 & \text{3/2} & - & - \\
 \text{32045.87(1)} & \text{0.2155(3)(7)} & 0.2325 & \text{5/2} & - & - \\
 \text{32060.03(1)} & \text{0.2107(3)(7)} & 0.2325 & \text{5/2} & - & - \\
 \text{32066.59(2)} & \text{0.2085(3)(7)} & 0.2325 & \text{5/2} & - & - \\
 \text{32071.04(2)} & \text{0.2061(5)(7)} & 0.2325 & \text{5/2} & - & - \\
 \text{32084.35(2)} & \text{0.2116(4)(7)} & 0.2325 & \text{3/2} & - & - \\
 \text{32127.05(2)} & \text{0.2117(5)(7)} & 0.2325 & \text{1/2} & - & $p2q=0.0090(18)$ \\
 \text{32177.07(3)} & \text{0.2130(6)(7)} & 0.2325 & \text{5/2} & - & - \\
 \text{32221.21(2)} & \text{0.2106(4)(7)} & 0.2325 & \text{1/2} & - & - \\
 \text{32266.95(2)} & \text{0.2105(5)(7)} & 0.2325 & \text{5/2} & - & - \\
 \text{32338.64(3)} & \text{0.2096(7)(7)} & 0.2325 & \text{5/2} & - & - \\
 \text{32369.53(1)} & \text{0.2152(3)(7)} & 0.2325 & \text{3/2} & - & - \\
 \text{32373.71(5)} & \text{0.2117(11)(7)} & 0.2325 & \text{5/2} & - & - \\
 \text{32460.91(3)} & \text{0.2096(4)(7)} & 0.2325 & \text{3/2} & - & - \\
 \text{32524.51(2)} & \text{0.2093(4)(7)} & 0.2325 & \text{1/2} & - & $p2q=0.0328(14)$ \\
 \text{32557.43(4)} & \text{0.2107(8)(7)} & 0.2325 & \text{1/2} & - & - \\
 \text{32560.96(2)} & \text{0.2146(5)} & 0.2332 & \text{5/2} & $?'-0''$ & - \\
 \text{32592.90(3)} & \text{0.2213(6)(7)} & 0.2325 & \text{3/2} & - & - \\
 \text{32642.68(3)} & \text{0.2061(6)(7)} & 0.2325 & \text{5/2} & - & - \\
 \text{32654.08(2)} & \text{0.2177(5)(7)} & 0.2325 & \text{3/2} & - & - \\
 \text{32705.70(2)} & \text{0.2096(4)(7)} & 0.2325 & \text{5/2} & - & - \\
 \text{32751.54(2)} & \text{0.2100(4)(7)} & 0.2325 & \text{3/2} & - & - \\
 \text{32854.45(2)} & \text{0.2161(3)(7)} & 0.2325 & \text{3/2} & - & - \\
 \text{32885.65(2)} & \text{0.2133(5)(7)} & 0.2325 & \text{5/2} & - & - \\
        \hline
        \end{tabular}
        \caption{Molecular parameters of all ThF transitions fitted. Units for $\nu_0$, $B'$, and $B''$ are in cm$^{-1}$.}
        \label{tab:all10}
    \end{table}
    
    \begin{table}[htb]
        \centering
        \begin{tabular}{| c c c c | c | c |}
        \hline
        $\nu_0$ & $B'$ & $B''$ & $\Omega'$ & $v'-v''$ & Additional comments \\
        \hline
 \text{32927.66(4)} & \text{0.2128(8)(7)} & 0.2325 & \text{1/2} & - & $p2q=0.0098(37)$ \\
 \text{32955.96(1)} & \text{0.2133(2)(7)} & 0.2325 & \text{3/2} & - & - \\
 \text{33013.58(3)} & \text{0.2248(10)(7)} & 0.2325 & \text{1/2} & - & - \\
 \text{33025.33(2)} & \text{0.2116(4)(7)} & 0.2325 & \text{1/2} & - & $p2q=0.0100(15)$ \\
 \text{33031.41(3)} & \text{0.2065(6)(7)} & 0.2325 & \text{5/2} & - & - \\
 \text{33033.36(3)} & \text{0.2119(6)(7)} & 0.2325 & \text{5/2} & - & - \\
 \text{33051.40(2)} & \text{0.2117(5)(7)} & 0.2325 & \text{1/2} & - & - \\
 \text{33072.15(1)} & \text{0.2121(3)(7)} & 0.2325 & \text{5/2} & - & - \\
 \text{33081.88(2)} & \text{0.2117(5)(7)} & 0.2325 & \text{5/2} & - & - \\
 \text{33120.12(2)} & \text{0.2131(5)(7)} & 0.2325 & \text{1/2} & - & $p2q=0.0295(17)$ \\
 \text{33129.32(4)} & \text{0.2128(9)(7)} & 0.2325 & \text{5/2} & - & - \\
 \text{33193.99(2)} & \text{0.2089(4)(7)} & 0.2325 & \text{5/2} & - & - \\
 \text{33256.47(3)} & \text{0.2131(5)(7)} & 0.2325 & \text{3/2} & - & - \\
 \text{33387.66(2)} & \text{0.2121(3)(7)} & 0.2325 & \text{3/2} & - & - \\
 \text{33413.67(4)} & \text{0.2116(6)(7)} & 0.2325 & \text{5/2} & - & - \\
 \text{33532.22(2)} & \text{0.2086(4)(7)} & 0.2325 & \text{1/2} & - & - \\
 \text{33567.61(3)} & \text{0.2128(5)(7)} & 0.2325 & \text{3/2} & - & - \\
 \text{33581.35(3)} & \text{0.2045(5)(7)} & 0.2325 & \text{5/2} & - & - \\
 \text{33632.10(2)} & \text{0.2061(5)(7)} & 0.2325 & \text{5/2} & - & - \\
 \text{33640.03(2)} & \text{0.2118(3)(7)} & 0.2325 & \text{5/2} & - & - \\
 \text{33781.49(2)} & \text{0.2116(4)(7)} & 0.2325 & \text{5/2} & - & - \\
 \text{33791.68(3)} & \text{0.2059(6)(7)} & 0.2325 & \text{5/2} & - & - \\
 \text{33815.27(1)} & \text{0.2165(3)(7)} & 0.2325 & \text{5/2} & - & - \\
 \text{34403.61(5)} & \text{0.2085(10)(7)} & 0.2325 & \text{1/2} & - & - \\
 \text{34405.80(2)} & \text{0.2130(5)(7)} & 0.2325 & \text{5/2} & - & - \\
        \hline
        \end{tabular}
        \caption{Molecular parameters of all ThF transitions fitted. Units for $\nu_0$, $B'$, and $B''$ are in cm$^{-1}$.}
        \label{tab:all11}
    \end{table}
    
    \begin{table}[htb]
        \centering
        \begin{tabular}{| c c c c | c | c |}
        \hline
        $\nu_0$ & $B'$ & $B''$ & $\Omega'$ & $v'-v''$ & Additional comments \\
        \hline
 \text{34638.18(2)} & \text{0.2132(4)(7)} & 0.2325 & \text{5/2} & - & - \\
 \text{36069.30(3)} & \text{0.2055(6)(7)} & 0.2325 & \text{5/2} & - & - \\
 \text{36083.10(3)} & \text{0.2142(7)(7)} & 0.2325 & \text{5/2} & - & - \\
 \text{36099.16(2)} & \text{0.2078(5)(7)} & 0.2325 & \text{1/2} & - & $p2q=0.0600(21)$ \\
 \text{36115.16(3)} & \text{0.2081(7)(7)} & 0.2325 & \text{5/2} & - & - \\
 \text{36168.34(2)} & \text{0.2139(4)(7)} & 0.2325 & \text{5/2} & - & - \\
 \text{36178.73(2)} & \text{0.2047(5)(7)} & 0.2325 & \text{5/2} & - & - \\
 \text{36197.34(2)} & \text{0.2137(5)} & 0.2318 & \text{5/2} & $?'-1''$ & - \\
 \text{36229.91(10)} & \text{0.2046(18)(7)} & 0.2325 & \text{5/2} & - & - \\
 \text{36239.10(3)} & \text{0.2084(5)(7)} & 0.2325 & \text{1/2} & - & - \\
 \text{36294.99(3)} & \text{0.2182(6)(7)} & 0.2325 & \text{5/2} & - & - \\
 \text{36463.34(2)} & \text{0.2105(4)(7)} & 0.2325 & \text{3/2} & - & - \\
 \text{36480.98(3)} & \text{0.2096(6)(7)} & 0.2325 & \text{1/2} & - & $p2q=0.0632(28)$ \\
 \text{36497.10(2)} & \text{0.2122(4)(7)} & 0.2325 & \text{3/2} & - & - \\
 \text{36654.63(2)} & \text{0.2056(4)(7)} & 0.2325 & \text{5/2} & - & - \\
 \text{36705.19(2)} & \text{0.2123(3)(7)} & 0.2325 & \text{5/2} & - & - \\
 \text{36760.16(2)} & \text{0.2149(6)(7)} & 0.2325 & \text{5/2} & - & - \\
 \text{36784.48(3)} & \text{0.2125(7)(7)} & 0.2325 & \text{5/2} & - & - \\
 \text{36794.17(2)} & \text{0.2148(5)} & 0.2332 & \text{5/2} & $?'-0''$ & - \\
 \text{36799.76(2)} & \text{0.2086(4)(7)} & 0.2325 & \text{3/2} & - & - \\
 \text{36845.42(2)} & \text{0.2094(4)(7)} & 0.2325 & \text{3/2} & - & - \\
 \text{36866.36(4)} & \text{0.2081(6)} & 0.2318 & \text{3/2} & $?'-1''$ & - \\
 \text{36878.58(3)} & \text{0.2092(5)} & 0.2318 & \text{3/2} & $?'-1''$ & - \\
 \text{36937.71(2)} & \text{0.2127(5)(7)} & 0.2325 & \text{5/2} & - & - \\
 \text{36949.82(10)} & \text{0.2031(21)(7)} & 0.2325 & \text{3/2} & - & - \\
        \hline
        \end{tabular}
        \caption{Molecular parameters of all ThF transitions fitted. Units for $\nu_0$, $B'$, and $B''$ are in cm$^{-1}$.}
        \label{tab:all12}
    \end{table}
    
    \begin{table}[htb]
        \centering
        \begin{tabular}{| c c c c | c | c |}
        \hline
        $\nu_0$ & $B'$ & $B''$ & $\Omega'$ & $v'-v''$ & Additional comments \\
        \hline
 \text{36958.77(3)} & \text{0.2102(7)(7)} & 0.2325 & \text{3/2} & - & - \\
 \text{36964.38(3)} & \text{0.2041(6)(7)} & 0.2325 & \text{3/2} & - & - \\
 \text{36999.78(2)} & \text{0.2096(3)(7)} & 0.2325 & \text{3/2} & - & - \\
 \text{37011.15(2)} & \text{0.2130(6)(7)} & 0.2325 & \text{3/2} & - & - \\
 \text{37017.36(2)} & \text{0.2128(5)(7)} & 0.2325 & \text{5/2} & - & - \\
 \text{37028.17(2)} & \text{0.2117(4)(7)} & 0.2325 & \text{1/2} & - & - \\
 \text{37072.99(3)} & \text{0.2072(7)(7)} & 0.2325 & \text{3/2} & - & - \\
 \text{37109.24(2)} & \text{0.2139(4)(7)} & 0.2325 & \text{5/2} & - & - \\
 \text{37216.01(3)} & \text{0.2093(5)(7)} & 0.2325 & \text{5/2} & - & - \\
 \text{37224.74(2)} & \text{0.2089(6)(7)} & 0.2325 & \text{5/2} & - & - \\
 \text{37236.55(2)} & \text{0.2122(4)(7)} & 0.2325 & \text{3/2} & - & - \\
 \text{37251.32(3)} & \text{0.2004(7)(7)} & 0.2325 & \text{5/2} & - & - \\
 \text{37261.75(1)} & \text{0.1951(3)(7)} & 0.2325 & \text{5/2} & - & - \\
 \text{37274.10(2)} & \text{0.2001(4)(7)} & 0.2325 & \text{5/2} & - & - \\
 \text{37316.83(2)} & \text{0.1947(5)(7)} & 0.2325 & \text{5/2} & - & - \\
 \text{37330.12(2)} & \text{0.2089(3)(7)} & 0.2325 & \text{3/2} & - & - \\
 \text{37347.77(2)} & \text{0.2112(4)(7)} & 0.2325 & \text{3/2} & - & - \\
 \text{37452.16(4)} & \text{0.1997(7)(7)} & 0.2325 & \text{3/2} & - & - \\
 \text{37457.58(3)} & \text{0.2062(5)(7)} & 0.2325 & \text{3/2} & - & - \\
 \text{37463.22(2)} & \text{0.2089(4)} & 0.2332 & \text{3/2} & $?'-0''$ & - \\
 \text{37475.41(1)} & \text{0.2094(3)} & 0.2332 & \text{3/2} & $?'-0''$ & - \\
 \text{37550.37(2)} & \text{0.2086(4)(7)} & 0.2325 & \text{1/2} & - & - \\
 \text{37597.65(2)} & \text{0.2129(4)(7)} & 0.2325 & \text{5/2} & - & - \\
 \text{37643.71(2)} & \text{0.2114(3)(7)} & 0.2325 & \text{3/2} & - & - \\
 \text{37694.56(3)} & \text{0.2120(5)(7)} & 0.2325 & \text{3/2} & - & - \\
        \hline
        \end{tabular}
        \caption{Molecular parameters of all ThF transitions fitted. Units for $\nu_0$, $B'$, and $B''$ are in cm$^{-1}$.}
        \label{tab:all13}
    \end{table}
    
    \begin{table}[htb]
        \centering
        \begin{tabular}{| c c c c | c | c |}
        \hline
        $\nu_0$ & $B'$ & $B''$ & $\Omega'$ & $v'-v''$ & Additional comments \\
        \hline
 \text{37754.70(2)} & \text{0.2125(4)(7)} & 0.2325 & \text{3/2} & - & - \\
 \text{37774.55(2)} & \text{0.2116(4)(7)} & 0.2325 & \text{3/2} & - & - \\
 \text{38495.94(4)} & \text{0.2095(6)(7)} & 0.2325 & \text{3/2} & - & - \\
 \text{38587.82(8)} & \text{0.2111(14)(7)} & 0.2325 & \text{3/2} & - & - \\
 \text{38601.46(4)} & \text{0.2076(6)(7)} & 0.2325 & \text{3/2} & - & - \\
 \text{38614.22(6)} & \text{0.2011(11)(7)} & 0.2325 & \text{3/2} & - & - \\
 \text{38762.68(5)} & \text{0.2101(9)(7)} & 0.2325 & \text{5/2} & - & - \\
 \text{38870.95(2)} & \text{0.2157(4)(7)} & 0.2325 & \text{5/2} & - & - \\
 \text{39091.15(2)} & \text{0.2060(4)(7)} & 0.2325 & \text{3/2} & - & - \\
 \text{39337.07(3)} & \text{0.2092(5)(7)} & 0.2325 & \text{3/2} & - & - \\
 \text{39771.88(3)} & \text{0.2092(5)(7)} & 0.2325 & \text{3/2} & - & - \\
 \text{39804.60(4)} & \text{0.2167(10)(7)} & 0.2325 & \text{5/2} & - & - \\
 \text{41298.58(4)} & \text{0.2084(8)(7)} & 0.2325 & \text{5/2} & - & - \\
 \text{41603.21(3)} & \text{0.2128(7)(7)} & 0.2325 & \text{5/2} & - & - \\
 \text{41983.31(3)} & \text{0.2057(7)(7)} & 0.2325 & \text{5/2} & - & - \\
 \text{42157.41(3)} & \text{0.2155(5)(7)} & 0.2325 & \text{1/2} & - & - \\
 \text{42577.45(3)} & \text{0.2089(7)(7)} & 0.2325 & \text{5/2} & - & - \\
 \text{42765.13(4)} & \text{0.2040(9)(7)} & 0.2325 & \text{5/2} & - & - \\
 \text{44577.57(3)} & \text{0.2006(6)(7)} & 0.2325 & \text{5/2} & - & - \\
        \hline
        \end{tabular}
        \caption{Molecular parameters of all ThF transitions fitted. Units for $\nu_0$, $B'$, and $B''$ are in cm$^{-1}$.}
        \label{tab:all14}
    \end{table}

\clearpage
\section{All ThO fitted transitions}
    \setcounter{table}{0}
    \setcounter{figure}{0}
    The molecular constants of all ThO transitions that we have fitted are shown in Tables \ref{tab:allThO1} to \ref{tab:allThO3}. We fix $B'' = 0.3313\text{cm}^{-1}$, i.e.,\ the average of the rotational constant of the $v''=0$ and $v''=1$ of the ThO ground electronic state, unless we assign that transition as a member of vibrational progression. In that case, we use the $B''$ and $B'$ of the relevant lower and upper states of the transitions. The values in the first pair of parentheses of $B'$ are fitting errors, and those in the second pair of parentheses are uncertainties from unknown initial vibrational states. States labelled with capital letters are found in previous work\cite{ThOgatterer1957molecular,ThOvon1970rotational,ThOwentink1972isoelectronic,ThOedvinsson1965g,ThOzare1973direct,ThOackermann1973high,ThOhildenbrand1974mass,NISTdatabase}. All fitting error bars are quoted to 90\% confidence in the fit.
    \begin{table}[htb]
        \centering
        \begin{tabular}{| c c c c | c | c |}
        \hline
        $\nu_0$ & $B'$ & $B''$ & $\Omega'$ & $v'-v''$ & Additional comments \\
        \hline
 \text{14489.99(1)} & \text{0.3213(3)} & 0.3320 & 1 & $0'-0''$ & C state \\
 \text{15055.28(4)} & \text{0.3206(9)} & 0.3307 & 1 & $0'-1''$ & D state \\
 \text{15255.31(6)} & \text{0.3194(12)} & 0.3307 & 1 & $?'-1''$ & - \\
 \text{15320.34(2)} & \text{0.3205(4)} & 0.3320 & 1 & $1'-0''$ & C state \\
 \text{15429.39(3)} & \text{0.3231(6)} & 0.3307 & 0 & $0'-1''$ & E state \\
 \text{15889.53(2)} & \text{0.3193(5)} & 0.3307 & 1 & $1'-1''$ & D state \\
 \text{15946.25(2)} & \text{0.3208(4)} & 0.3320 & 1 & $0'-0''$ & D state \\
 \text{16146.31(3)} & \text{0.3191(7)} & 0.3320 & 1 & $?'-0''$ & - \\
 \text{16254.01(2)} & \text{0.3216(5)} & 0.3307 & 0 & $1'-1''$ & E state \\
 \text{16320.38(2)} & \text{0.3222(4)} & 0.3320 & 0 & $0'-0''$ & E state \\
 \text{18337.55(2)} & \text{0.3212(5)} & 0.3320 & 0 & $0'-0''$ & F state \\
 \text{19094.78(2)} & \text{0.3243(4)} & 0.3320 & 0 & $1'-0''$ & F state \\
 \text{19354.70(2)} & \text{0.3270(8)(7)} & 0.3313 & 1 & - & - \\
 \text{19445.97(2)} & \text{0.3262(8)} & 0.3307 & 1 & $1'-1''$ & I state \\
 \text{19539.09(1)} & \text{0.3277(3)} & 0.3320 & 1 & $0'-0''$ & I state \\
 \text{20336.94(2)} & \text{0.3269(5)} & 0.3320 & 1 & $1'-0''$ & I state \\
        \hline
        \end{tabular}
        \caption{Molecular parameters of all ThO transitions fitted. Units for $\nu_0$, $B'$, and $B''$ are in cm$^{-1}$.}
        \label{tab:allThO1}
    \end{table}
    
    \begin{table}[htb]
        \centering
        \begin{tabular}{| c c c c | c | c |}
        \hline
        $\nu_0$ & $B'$ & $B''$ & $\Omega'$ & $v'-v''$ & Additional comments \\
        \hline
 \text{27648.53(2)} & \text{0.3122(4)(7)} & 0.3313 & 1 & - & - \\
 \text{27719.27(1)} & \text{0.3150(2)(7)} & 0.3313 & 1 & - & - \\
 \text{28028.83(2)} & \text{0.3176(3)(7)} & 0.3313 & 0 & - & - \\
 \text{28182.04(4)} & \text{0.3173(7)} & 0.3307 & 1 & $0'-1''$ & $\Omega=1$ [29.07] \\
 \text{28578.20(5)} & \text{0.3270(10)(7)} & 0.3313 & 0 & - & - \\
 \text{29072.94(2)} & \text{0.3176(3)} & 0.3320 & 1 & $0'-0''$ & $\Omega=1$ [29.07] \\
 \text{29401.72(2)} & \text{0.3229(3)(7)} & 0.3313 & 1 & - & - \\
 \text{29867.81(1)} & \text{0.3168(2)} & 0.3320 & 1 & $1'-0''$ & $\Omega=1$ [29.07] \\
 \text{30217.45(2)} & \text{0.3205(4)(7)} & 0.3313 & 1 & - & - \\
 \text{30243.81(5)} & \text{0.3234(9)(7)} & 0.3313 & 1 & - & - \\
 \text{30313.01(1)} & \text{0.3257(2)(7)} & 0.3313 & 1 & - & - \\
 \text{30646.33(2)} & \text{0.3249(5)} & 0.3307 & 0 & $?'-1''$ & - \\
 \text{30717.79(2)} & \text{0.3228(6)(7)} & 0.3213 & 0 & - & - \\
 \text{30895.14(3)} & \text{0.3222(6)(7)} & 0.3313 & 1 & - & - \\
 \text{30959.95(2)} & \text{0.3268(4)(7)} & 0.3313 & 0 & - & - \\
 \text{31537.28(3)} & \text{0.3250(10)} & 0.3320 & 0 & $?'-0''$ & - \\
 \text{32777.37(1)} & \text{0.3157(3)} & 0.3307 & 0 & $?'-1''$ & - \\
 \text{32866.31(2)} & \text{0.3170(3)(7)} & 0.3313 & 0 & - & - \\
 \text{33106.17(1)} & \text{0.3252(2)} & 0.3307 & 1 & $?'-1''$ & - \\
 \text{33303.79(2)} & \text{0.3214(3)(7)} & 0.3313 & 1 & - & - \\
 \text{33480.52(2)} & \text{0.3139(4)(7)} & 0.3313 & 0 & - & - \\
 \text{33574.54(2)} & \text{0.3149(3)(7)} & 0.3313 & 0 & - & - \\
 \text{33629.12(2)} & \text{0.3241(4)(7)} & 0.3313 & 1 & - & - \\
 \text{33668.33(1)} & \text{0.3161(3)} & 0.3320 & 0 & $?'-0''$ & - \\
 \text{33725.15(2)} & \text{0.3241(4)(7)} & 0.3313 & 1 & - & - \\
 \text{33822.44(2)} & \text{0.3232(5)(7)} & 0.3313 & 0 & - & - \\
 \text{33872.48(2)} & \text{0.3230(4)(7)} & 0.3313 & 1 & - & - \\
 \text{33917.09(2)} & \text{0.3217(4)(7)} & 0.3313 & 0 & - & - \\
 \text{33947.12(1)} & \text{0.3345(5)(7)} & 0.3313 & 1 & - & - \\
        \hline
        \end{tabular}
        \caption{Molecular parameters of all ThO transitions fitted. Units for $\nu_0$, $B'$, and $B''$ are in cm$^{-1}$.}
        \label{tab:allThO2}
    \end{table}
    
    \begin{table}[htb]
        \centering
        \begin{tabular}{| c c c c | c | c |}
        \hline
        $\nu_0$ & $B'$ & $B''$ & $\Omega'$ & $v'-v''$ & Additional comments \\
        \hline
 \text{33997.20(3)} & \text{0.3251(5)} & 0.3320 & 1 & $?'-0''$ & - \\
 \text{34030.28(2)} & \text{0.3221(3)(7)} & 0.3313 & 0 & - & - \\
 \text{34094.70(1)} & \text{0.3196(3)(7)} & 0.3313 & 1 & - & - \\
 \text{34115.25(5)} & \text{0.3240(11)(7)} & 0.3313 & 1 & - & - \\
 \text{34132.34(1)} & \text{0.3219(2)(7)} & 0.3313 & 0 & - & - \\
 \text{34154.62(3)} & \text{0.3213(5)(7)} & 0.3313 & 1 & - & - \\
 \text{34165.25(3)} & \text{0.3098(7)(7)} & 0.3313 & 0 & - & - \\
 \text{34197.18(2)} & \text{0.3215(2)(7)} & 0.3313 & 1 & - & - \\
 \text{34216.46(3)} & \text{0.3232(5)(7)} & 0.3313 & 1 & - & - \\
 \text{34267.21(4)} & \text{0.3124(7)(7)} & 0.3313 & 0 & - & - \\
 \text{34314.57(2)} & \text{0.3151(4)(7)} & 0.3313 & 1 & - & - \\
 \text{34346.74(2)} & \text{0.3106(3)(7)} & 0.3313 & 1 & - & - \\
 \text{34352.33(3)} & \text{0.3433(6)(7)} & 0.3313 & 1 & - & - \\
 \text{34365.48(8)} & \text{0.3189(18)(7)} & 0.3313 & 1 & - & - \\
 \text{34385.28(1)} & \text{0.3213(3)(7)} & 0.3313 & 1 & - & - \\
 \text{34510.53(4)} & \text{0.3221(10)(7)} & 0.3313 & 1 & - & - \\
 \text{34547.96(3)} & \text{0.3219(6)(7)} & 0.3313 & 1 & - & - \\
 \text{34611.28(2)} & \text{0.3248(6)(7)} & 0.3313 & 1 & - & - \\
 \text{35580.50(2)} & \text{0.3208(4)(7)} & 0.3313 & 0 & - & - \\
 \text{35684.62(4)} & \text{0.3207(11)(7)} & 0.3313 & 0 & - & - \\
 \text{35723.63(2)} & \text{0.3305(6)(7)} & 0.3313 & 0 & - & - \\
 \text{35807.37(2)} & \text{0.3214(6)(7)} & 0.3313 & 0 & - & - \\
 \text{35892.33(1)} & \text{0.3233(3)(7)} & 0.3313 & 1 & - & - \\
 \text{35922.21(4)} & \text{0.3205(7)(7)} & 0.3313 & 1 & - & - \\
 \text{35993.67(2)} & \text{0.3192(3)(7)} & 0.3313 & 1 & - & - \\
 \text{37807.94(2)} & \text{0.3159(4)} & 0.3307 & 1 & $?'-1''$ & - \\
 \text{37890.03(3)} & \text{0.3159(6)(7)} & 0.3313 & 1 & - & - \\
        \hline
        \end{tabular}
        \caption{Molecular parameters of all ThO transitions fitted. Units for $\nu_0$, $B'$, and $B''$ are in cm$^{-1}$.}
        \label{tab:allThO3}
    \end{table}
    
    \begin{table}[htb]
        \centering
        \begin{tabular}{| c c c c | c | c |}
        \hline
        $\nu_0$ & $B'$ & $B''$ & $\Omega'$ & $v'-v''$ & Additional comments \\
        \hline
 \text{38026.22(1)} & \text{0.3193(2)(7)} & 0.3313 & 1 & - & - \\
 \text{38156.51(2)} & \text{0.3174(4)(7)} & 0.3313 & 1 & - & - \\
 \text{38193.77(2)} & \text{0.3094(4)(7)} & 0.3313 & 1 & - & - \\
 \text{38698.91(2)} & \text{0.3148(5)} & 0.3320 & 1 & $?'-0''$ & - \\
 \text{39548.79(4)} & \text{0.3015(8)(7)} & 0.3313 & 1 & - & - \\
 \text{39595.61(2)} & \text{0.3171(3)(7)} & 0.3313 & 1 & - & - \\
 \text{39744.25(3)} & \text{0.3071(7)(7)} & 0.3313 & 1 & - & - \\
 \text{39749.19(4)} & \text{0.3157(9)(7)} & 0.3313 & 1 & - & - \\
 \text{39895.61(2)} & \text{0.3010(5)(7)} & 0.3313 & 0 & - & - \\
 \text{40193.90(2)} & \text{0.3100(5)(7)} & 0.3313 & 1 & - & - \\
 \text{40619.87(4)} & \text{0.3065(10)(7)} & 0.3313 & 0 & - & - \\
        \hline
        \end{tabular}
        \caption{Molecular parameters of all ThO transitions fitted. Units for $\nu_0$, $B'$, and $B''$ are in cm$^{-1}$.}
        \label{tab:allThO4}
    \end{table}

\end{document}